\documentclass[preprint]{aastex}

\slugcomment{Accepted to the Astronomical Journal, 13 May 2005}

\shorttitle{Properties of Protostars From Resolved Near-IR Photospheric Lines}
\shortauthors{Doppmann et al.}

\begin{document}


\title{The Physical Natures of Class I and Flat-Spectrum Protostellar
Photospheres: A Near-Infrared Spectroscopic Study\footnote{Data
presented herein were obtained at the W.M. Keck Observatory from
telescope time allocated to the National Aeronautics and Space
Administration through the agency's scientific partnership with the
California Institute of Technology and the University of California.
The Observatory was made possible by the generous financial support of
the W.M. Keck Foundation.}}


\author{G. W. Doppmann\altaffilmark{2,3},
T. P. Greene\altaffilmark{2}, K. R. Covey\altaffilmark{4}, and
C. J. Lada\altaffilmark{5}}

\email{doppmann@gemini.edu}
\email{thomas.p.greene@nasa.gov}
\email{covey@astro.washington.edu}
\email{clada@cfa.harvard.edu}


\altaffiltext{2}{NASA Ames Research Center, Mail Stop 245-6, Moffett
Field, CA 98035-1000}

\altaffiltext{3}{Present address: Gemini Observatory, Southern
Operations Center, Association of Universities for Research in
Astronomy, Inc., Casilla 603, La Serena, Chile}

\altaffiltext{4}{University of Washington, Department of Astronomy,
Box 351580, Seattle, WA 98195}

\altaffiltext{5}{Harvard-Smithsonian Center for Astrophysics, 60
Garden Street, Cambridge, MA 02138}


\begin{abstract}

We present high resolution (R $\simeq$ 18,000), high signal-to-noise,
2 $\mu$m spectra of 52 infrared-selected Class I and flat-spectrum
young stellar objects in the Taurus-Auriga, $\rho$ Ophiuchi, Serpens,
Perseus, and Corona Australis dark clouds.  We detect key absorption
lines in 41 objects and fit synthetic spectra generated from pre-main
sequence models to deduce the effective temperatures, surface
gravities, near-infrared veilings, rotation velocities, and radial
velocities of each of these 41 sources.  We find these objects to span
ranges in effective temperature, surface gravity, and stellar
luminosity which appear similar to those of late spectral-type Class
II sources and classical T-Tauri stars.  However, due to significant
but uncertain corrections for scattering and extinction, the derived
luminosities for the embedded protostellar objects must be regarded as
being highly uncertain.  We determine that the mean 2 $\mu$m veiling
of Class I and flat-spectrum objects is significantly higher than that
of Class II objects in the same region where both types of objects are
extensively observed ($\rho$ Oph).  We find a significant fraction of
our protostellar sample also exhibits emission lines.  Twenty-three
objects show H$_2$ emission, usually indicative of the presence of
energetic outflows.  Thirty-four sources show HI Br~$\gamma$ emission
and a number of these exhibit profile asymmetries consistent with
in-fall.  Eight sources show significant $\Delta v = 2$ CO emission
suggestive of emission from a circumstellar disk.  Overall, these
observations indicate that Class I and flat-spectrum objects are
self-embedded protostars undergoing significant mass accretion,
although the objects appear to span a broad range of mass accretion
activity.

\end{abstract}





\section{Introduction}

The quest to identify and understand protostars has long been a key
but stubborn problem for star formation research \citep{spitzer1948,
larson1969, wynnwilliams1982, shu1987}.  The primary factor hindering
the investigation of protostellar objects has been the fact that such
sources are deeply embedded in dusty molecular gas and are heavily
extincted.  As a result most of the luminosity generated by a
protostar is absorbed and scattered by dust and radiated in the
infrared, rendering the object very difficult, if not impossible, to
detect at optical wavelengths.  Consequently, progress toward
achieving a physical understanding of the nature of protostars had to
wait for the development of observing capabilities at longer
wavelengths.

Significant progress in understanding the physical nature and
evolutionary status of protostars has been achieved from measurements
of the broadband photometric infrared spectral energy distributions
(SEDs) of low-luminosity young stellar objects (YSOs) in star forming
molecular clouds \citep{wilking1983, lada1987, adams1987, andre1993}.
These studies established that the shape of the SED of a YSO is linked
to its evolutionary state.  Class I (and Class 0) sources were
characterized by strong infrared excesses and steeply rising SEDs from
near- to far-infrared wavelengths.  These sources were found to be
distinctly different from Class II sources.  These latter objects were
characterized by smaller infrared excesses and SEDs that decreased
from near- to far-infrared wavelengths and could be
well-explained by models of pre-main sequence (PMS) stars surrounded
by circumstellar disks (i.e.  classical T-Tauri stars).  Class II
sources, in turn, were differentiated from Class III sources which
displayed no infrared excess and whose SEDs could be explained by
normal stellar photospheres of classical PMS stars.  Class I SEDs were
found to be consistent with, and well-fit by, theoretical models of
protostellar envelopes.  Moreover, other lines of evidence, such as
the intimate association of Class I (and 0) sources with molecular
outflows, molecular hydrogen emission, FU Ori phenomenon, etc.
strongly suggested an extreme youth and possibly protostellar nature
for these objects.

Little is known about the underlying stellar component of protostars
since these objects are typically undetectable in visible light
rendering the traditional methods of optical spectroscopy that are
used to measure stellar physical properties not generally applicable.
However advances in technology over the past dozen years have enabled
spectroscopic observations of Class I sources at near-infrared
wavelengths. In a pioneering paper, \citet{casali1992} obtained low
resolution (R~$\sim$~300) spectra at 2 $\mu$m of a sample of ten
bright Class I and Class II sources in Ophiuchus and found Class I
sources to exhibit a featureless continuum, unlike Class II sources
which displayed photospheric absorption lines. This result was
confirmed by a subsequent low resolution (R $\sim$ 500) near-infrared
spectroscopic survey of a significantly larger sample of $\sim$100 low
luminosity YSOs in Ophiuchus and Taurus \citep{greene1996}. The
absence of photospheric absorption lines in Class I sources was
interpreted as being the result of high veiling due to strong infrared
excess emission produced by disk accretion and reprocessing in
circumstellar envelopes. This provided strong supporting evidence for
a protostellar nature for Class I objects.  However, the high veilings
of Class I sources indicated that very high signal-to-noise and high
spectral resolution observations would be required to detect
photospheric features and measure the physical parameters of the
embryonic stellar core at the heart of a protostellar object.

\citet{greene1997, greene2000} obtained the first such deep and high
resolution (R $\sim$ 20,000) infrared spectra and detected
photospheric absorption features in a sample of several flat-spectrum
protostars. These embedded sources exhibit SEDs with shapes between
those of Class I and II objects and are believed to be in transition
between the Class I and II evolutionary states. Flat-spectrum sources
are likely to be extremely young and can be considered the least
obscured candidate protostars. These objects were found to have late
(M0 or later) spectral-types confirming their nature as low mass
stellar objects and to be relatively fast rotators compared to Class
II sources, an unexpected result. \citet{greene2000} also obtained
spectra of five Class I sources, and all presented featureless spectra
indicating that even more sensitive infrared observations would be
necessary to detect the photospheres of these objects. Using the 10
meter Keck telescope, \citet{greene2002} were able to obtain yet
deeper observations and detect the first photospheric absorption lines
in a heavily veiled accreting protostellar source, YLW~15.  These
observations enabled determinations of the central source's
spectral-type, veiling, stellar luminosity and allowed its placement
on the Hertzsprung-Russell (H-R) diagram.

Although most Class I sources are deeply buried in dusty envelopes,
visible light scattered off cavities in these envelopes can in some
cases be detected at optical wavelengths with large ground-based
telescopes. \citet{kenyon1998} performed an I-band spectroscopic
survey of 13 Class I sources in Taurus that were optically
visible. They were able to detect photospheric features in 5 of these
sources and found them to display late (M) spectral-types and to lie
in the same general region of the H-R diagram as Class II sources. The
optically-selected sample of Taurus Class I sources also did exhibit
stronger and more frequent forbidden-line emission than Class II
sources, suggesting higher mass outflow and accretion rates. A recent
higher resolution, more sensitive optical spectroscopic survey of
optically detected Class I sources in Taurus by \citet{white2004}
produced spectral-types of 11 objects and generally confirmed the
trends observed by \citet{kenyon1998}, although the \citet{white2004}
observations suggest similar accretion and rotation rates for Class I
and II objects in their sample.  These observations suggest that
optically detectable Class I sources may be in a more advanced state
of evolution than the more common optically invisible Class I
protostars which are more deeply embedded.

Sensitive high resolution infrared and optical spectroscopy of
flat-spectrum and Class I sources represent a significant advance in
protostellar studies and clearly have demonstrated the utility and
potential of such observations for investigating protostellar
evolution. A number of interesting issues have been raised by these
observations. Do the optically detectable Class I sources represent
the end stages of protostellar evolution? Are their physical
properties similar to those of the optically invisible Class I objects
or is there a significant range in evolutionary states among Class I
objects from accretion active to accretion anemic objects?

To better define the physical properties and nature of Class I sources
and achieve a more complete understanding of protostellar evolution we
have conducted the first extensive and systematic near-infrared
spectral survey of deeply embedded Class I and flat-spectrum objects
carried out at high resolution and with the high sensitivity afforded
by a 10 meter class telescope.  We were able to obtain high quality
spectra for more than 50 Class I and flat-spectrum sources in three
nearby regions of active star formation.  Photospheric absorption
features were detected with good signal-to-noise in the vast majority
of these sources. In order to derive the physical parameters that
characterize these young stellar objects we have employed synthetic
model atmospheres following the techniques successfully developed by
\citet{doppmann2003a} (hereafter DJ03) in their recent study of
infrared Class II sources in Ophiuchus \citep{doppmann2003b}.
 
In this paper we present the basic observational results of the
survey. In $\S$ 2 we describe the instrumentation and observing
procedures used to obtain our data and discuss our data reduction
methods. In $\S$ 3 we present the results and analysis of our
observations. In $\S$ 4 we discuss the implications for the nature of
Class I and flat-spectrum sources in the context of protostellar
evolution and in $\S$ 5 we summarize the results and conclusions of
this contribution.  This paper is the first in a series of three
papers dealing with the results of our survey. Two subsequent papers
will respectively deal with the analysis of the rotational properties
\citep{covey2005a} and radial velocities \citep{covey2005b} of this
sample.

\section{Observations and Data Reduction}

\subsection{Sample Selection}

YSOs were selected from nearby star forming regions of $\rho$
Oph, Taurus, Serpens, Perseus, and Corona Australis using the
following criteria: (1) objects with rising or flat spectral energy
distributions across 2 -- $\ge$10 $\mu$m, indicative of excess
emission at infrared wavelengths ($\alpha \equiv
dlog(\lambda~F_{\lambda}) / dlog(\lambda) > -0.3$); and (2) except for
a few fainter sources in Ophiuchus, our targets were generally K $\le$
11, in order to obtain high signal-to-noise spectra (S/N $>$ 100) in a
reasonable amount of time ($\le$ 120 minutes) at the telescope.  In
total, our sample was comprised of 52 flat-spectrum or Class I
protostars: 2 in Perseus, 13 in Taurus, 20 in Ophiuchus, 13 in
Serpens, and 4 in Corona Australis (Table \ref{tbl-1}).  These sources
were selected from
\citet{wilking1989,wilking1992,ladd1993,greene1994,kenyon1995,kaas2004}.
In addition, 28 late-type giant and dwarf MK spectral standards were
observed for calibration of our spectral synthesis models.

\subsection{Spectroscopic Observations}

Near-IR spectra of the protostellar sample and MK spectral standards
were acquired on 2000 May 29 -- 30, 2001 July 7 -- 10, 2001 November 4
-- 6, and 2003 June 19 -- 21 UT.  All data were acquired with the 10-m
Keck II telescope on Mauna Kea, Hawaii, using the NIRSPEC multi-order
cryogenic echelle facility spectrograph \citep{mclean1998}.  Spectra
were acquired with a 0\farcs58 (4 pixel) wide slit, providing
spectroscopic resolution $R \equiv \lambda / \delta \lambda$ = 18,000
(16.7 km~s$^{-1}$).  The plate scale was 0\farcs20 pixel$^{-1}$ along
the 12$\arcsec$ slit length, and the seeing was typically 0\farcs5 --
0\farcs6. The NIRSPEC gratings were oriented to allow orders
containing the 2.1066 $\mu$m Mg and 2.1099 $\mu$m Al lines, the 2.1661
$\mu$m HI Br~$\gamma$ line, the 2.2062 and 2.2090 $\mu$m Na lines, and
the 2.2935 $\mu$m CO bandhead regions to fall onto the instrument's
1024 $\times$ 1024 pixel InSb detector array. The instrument's
NIRSPEC-7 blocking filter was used to image these orders on the
detector. NIRSPEC was configured to acquire simultaneously multiple
cross-dispersed echelle orders 31 -- 36 (2.08 -- 2.45 $\mu$m,
non-continuous) for most objects, but some objects were observed in
orders 32 -- 38 (1.97 -- 2.38 $\mu$m, non-continuous). Each order had
an observed spectral range $\Delta \lambda \simeq \lambda / 67$
($\Delta v \simeq$ 4450 km~s$^{-1}$).

The internal instrument rotator was used to maintain a fixed position
angle on the sky when observing on 2000 May 29 -- 30.  Unfortunately
vibration from the rotator caused slight changes in the echelle
grating angle during those nights, but we removed the resultant
wavelength shifts during the data reduction process. The slit was held
physically stationary and thus allowed to rotate on the sky (as the
non-equatorially-mounted telescope tracked) when observing at all
other times.  Data were acquired in pairs of exposures of durations
from less than 1 second (for giant MK standards) to up to 900 s (for
the protostars) each, with the telescope nodded $6\arcsec$ along the
slit between frames so that object spectra were acquired in all
exposures.  Early-type (B7 -- A2) dwarfs were observed for telluric
correction of the protostellar and MK standard stellar spectra.  The
telescope was automatically guided with frequent images from the
NIRSPEC internal SCAM IR camera during all exposures of more than
several seconds duration.  Spectra of the internal NIRSPEC continuum
lamp were taken for flat fields, and exposures of the Ar, Ne, Kr, and
Xe lamps were used for wavelength calibrations. The observation dates,
total integration times, signal-to-noise, and other parameters of the
observed objects are given in Table \ref{tbl-1}.

\subsection{Data Reduction}

All data were reduced following standard procedures with IRAF packages
\citep{massey1992,massey1997}.  Bad pixels and cosmic rays were
removed with the COSMICRAY task -- images were cleaned once to remove
cosmic rays and bad pixels, then inverted and cleaned again to remove
cold pixels before being returned to positive flux.  Frames were then
flatfield divided and sky subtracted before spectra were extracted
with the APALL task.  Extracted spectra were wavelength calibrated
using 4th order fits to lines in arc lamp exposures, and all spectra
of objects at the same slit position and similar airmasses were
coadded.  Instrument vibration during the May 2000 observing run
caused the echelle grating to shift slightly between lamp and object
spectra exposures.  Corrections for the shift of the wavelength
solution between objects were applied by using the FXCOR task to find
the pixel shift between object spectra and telluric standards observed
closely after lamp spectra.  This appears to have improved the
wavelength solutions of the May 2000 data considerably, but radial
velocities derived for objects observed during May 2000 will be
significantly more uncertain than those objects observed on later
runs.

Instrumental and telluric features were removed by dividing wavelength
calibrated object spectra by spectra of early type stars observed at
similar airmass and slit position.  Spectral orders containing the HI
Br~$\gamma$ line (2.1661 $\mu$m) were corrected for telluric
absorption with the use of the XTELLCOR IDL routine developed by
\citet{vacca2003}.  This routine creates and then removes a pure
telluric spectrum generated by dividing an observed early-type
stellar spectrum by a synthetic model of Vega (modified to include
line broadening).  Combined spectra were then produced by summing the
spectra of both slit positions for each object.  Wavelength shifts
have not been applied to correct for Earth and solar system motions;
rather, radial velocities have been measured and then converted to
radial velocities with respect to the local standard of rest.
Finally, the spectra of each wavelength sub-interval used in our
analysis (see $\S$\ref{sec-diaglines}), was flattened and continuum
normalized to one by linear function, so as to be directly comparable
to the intrinsically flat continua of our synthesis models.

\begin{figure}
\figurenum{1}
\plotone{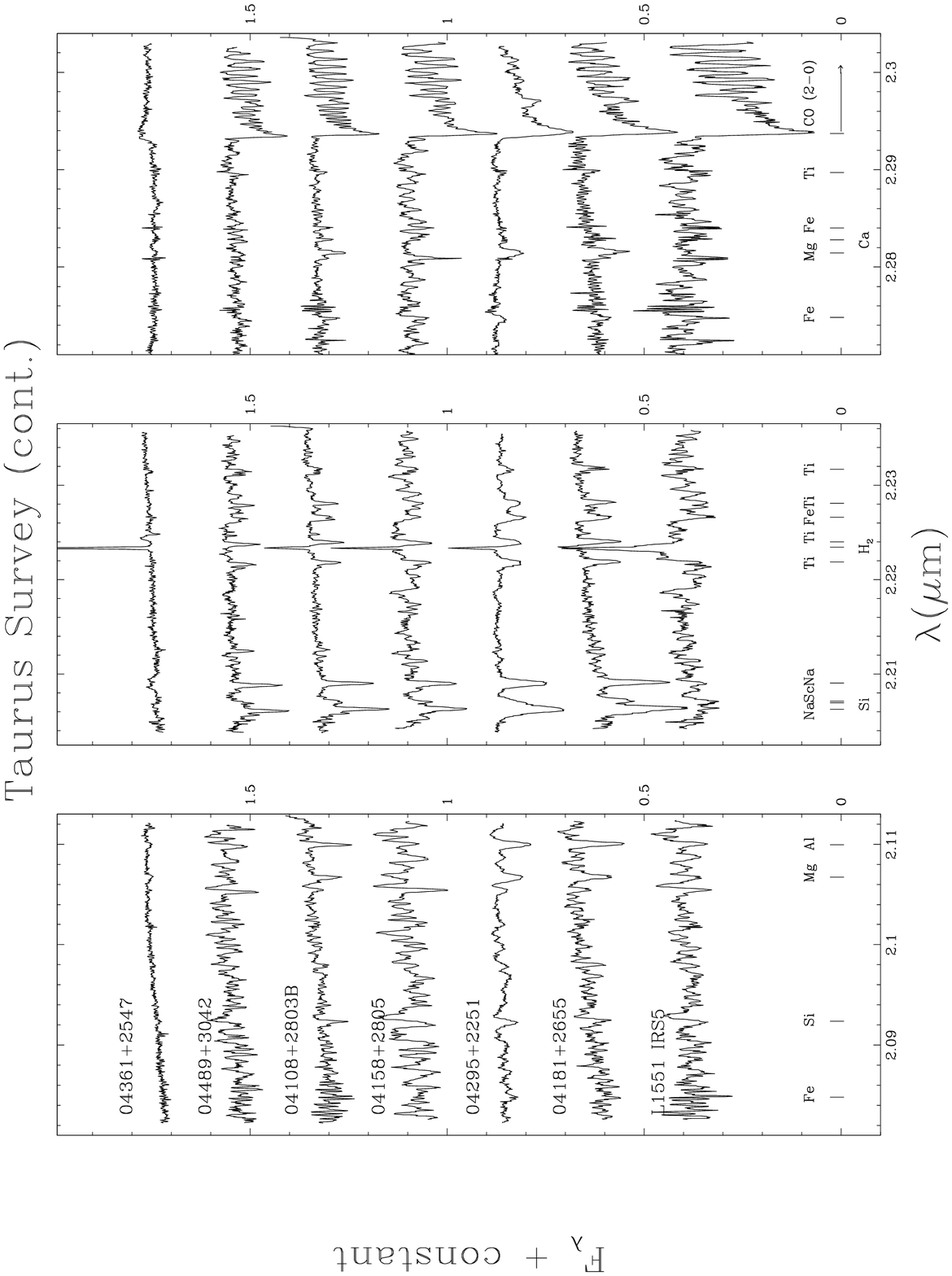}
\caption[Taurus Spectra2]
{\label{fig-all.spectra2}}

\end{figure}

\begin{figure}
\figurenum{2}
\plotone{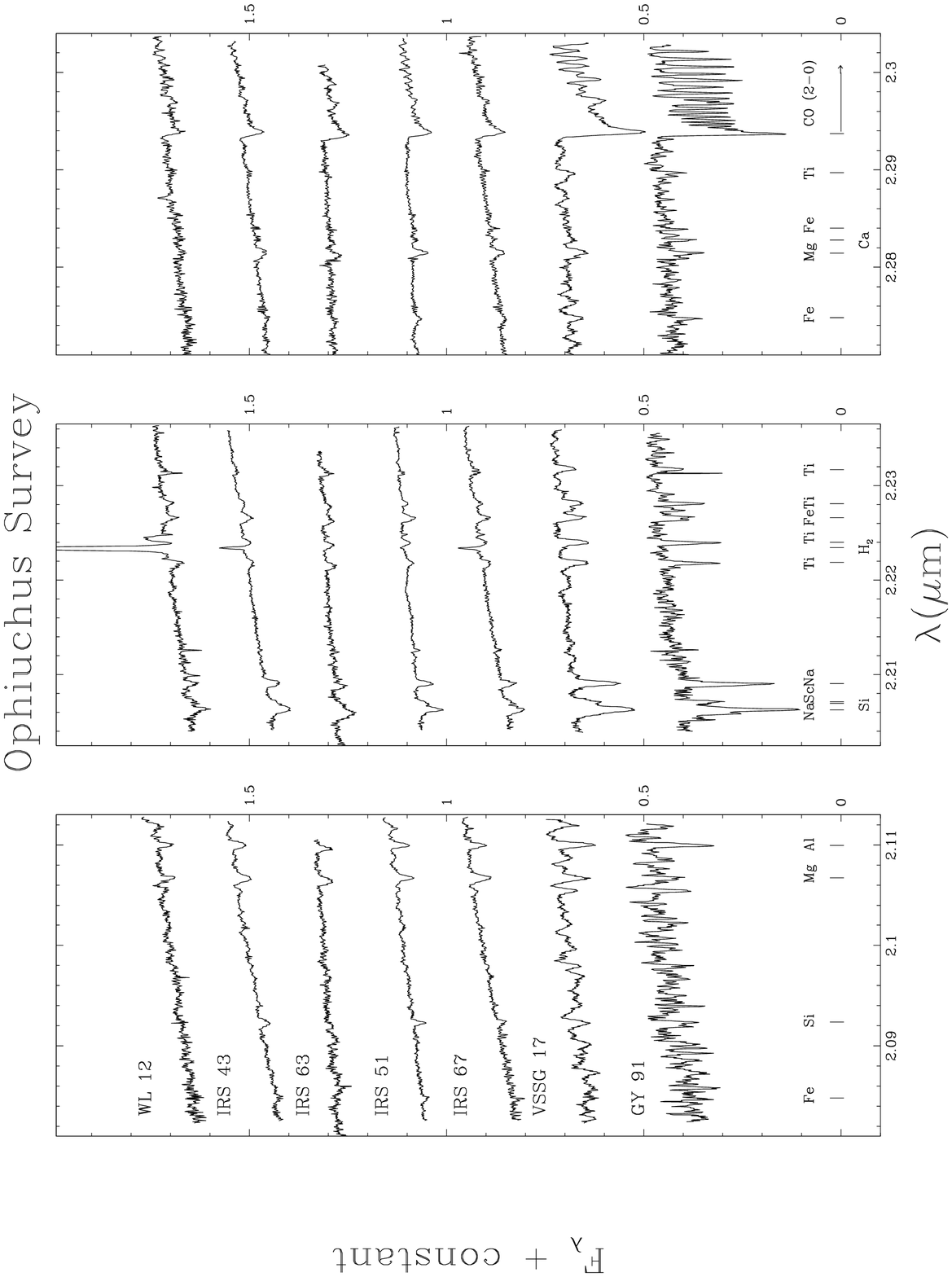}
\caption[Ophiuchus Spectra1]
{\label{fig-all.spectra3}
High resolution (R $\simeq$ 18,000) spectra of protostars in Ophiuchus
are shown in three echelle orders, which contain the photospheric
absorption line regions (Mg/Al, Na, and $^{12}$CO intervals, bottom to
top, respectively) that we use to derive physical properties using
spectral fits to synthetic templates.  At this resolution, the
intrinsic line widths of K-band photospheric lines are fully resolved
allowing precise fits to their line shapes.}

\end{figure}

\begin{figure}
\figurenum{2}
\plotone{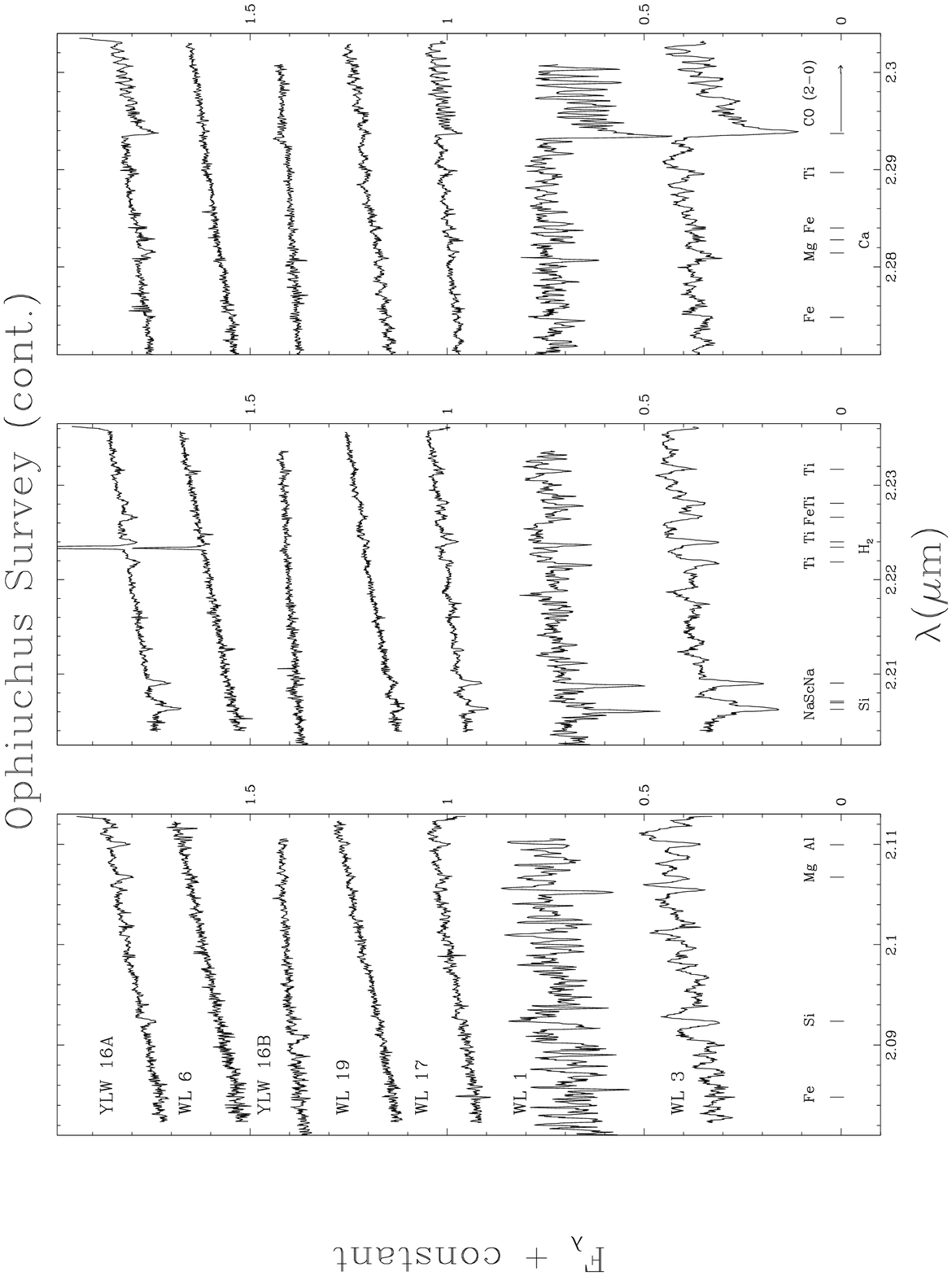}
\caption[Ophiuchus Spectra2]
{\label{fig-all.spectra4}}

\end{figure}

\begin{figure}
\figurenum{2}
\plotone{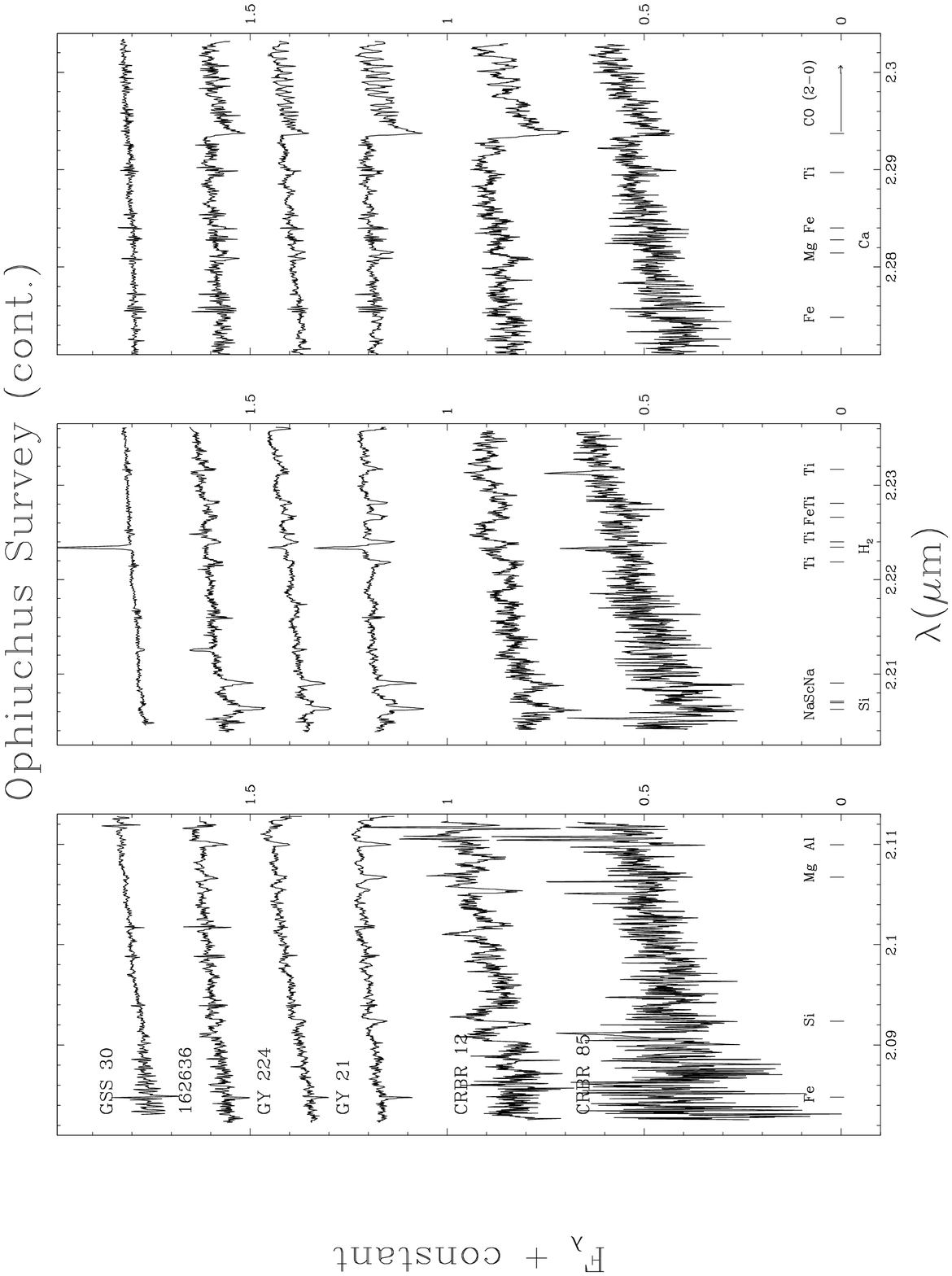}
\caption[Ophiuchus Spectra3]
{\label{fig-all.spectra5}}

\end{figure}

\begin{figure}
\figurenum{3}
\plotone{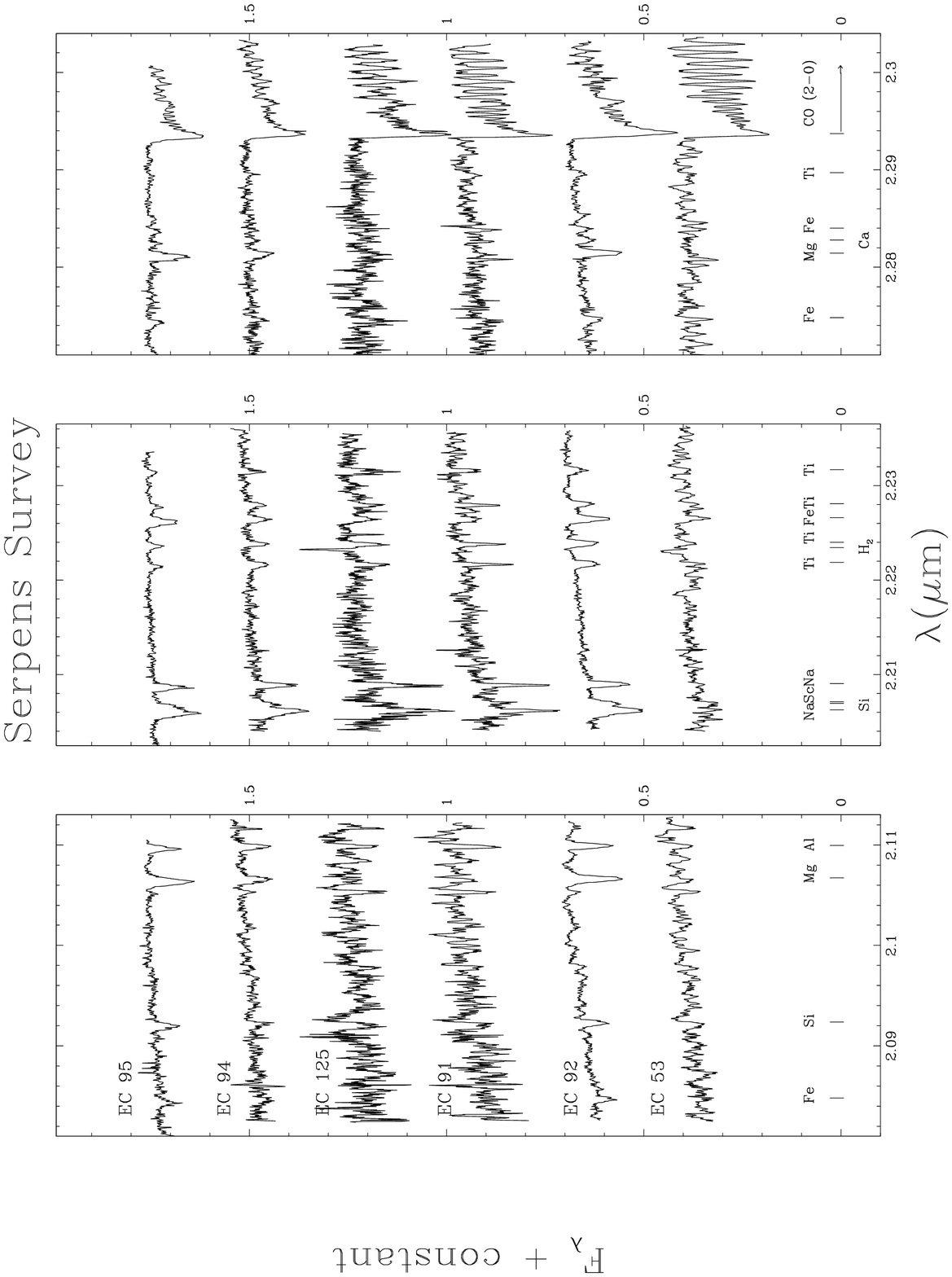}
\caption[Serpens Spectra1]
{\label{fig-all.spectra6}
High resolution (R $\simeq$ 18,000) spectra of protostars in Serpens
are shown in three echelle orders, which contain the photospheric
absorption line regions (Mg/Al, Na, and $^{12}$CO intervals, bottom to
top, respectively) that we use to derive physical properties using
spectral fits to synthetic templates.  At this resolution, the
intrinsic line widths of K-band photospheric lines are fully resolved
allowing precise fits to their line shapes.}

\end{figure}

\begin{figure}
\figurenum{3}
\plotone{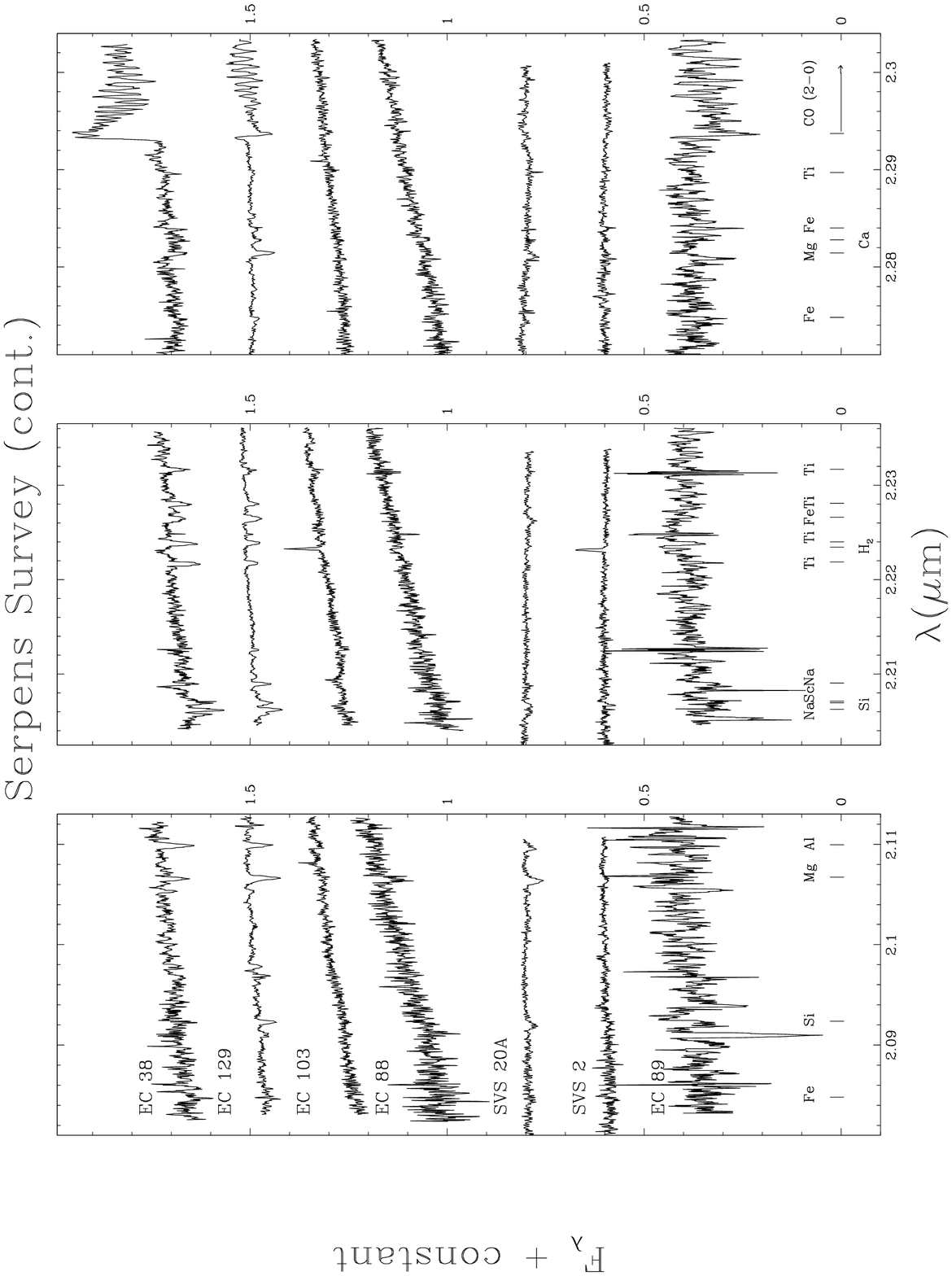}
\caption[Serpens Spectra2]
{\label{fig-all.spectra7}}

\end{figure}

\begin{figure}
\figurenum{4}
\plotone{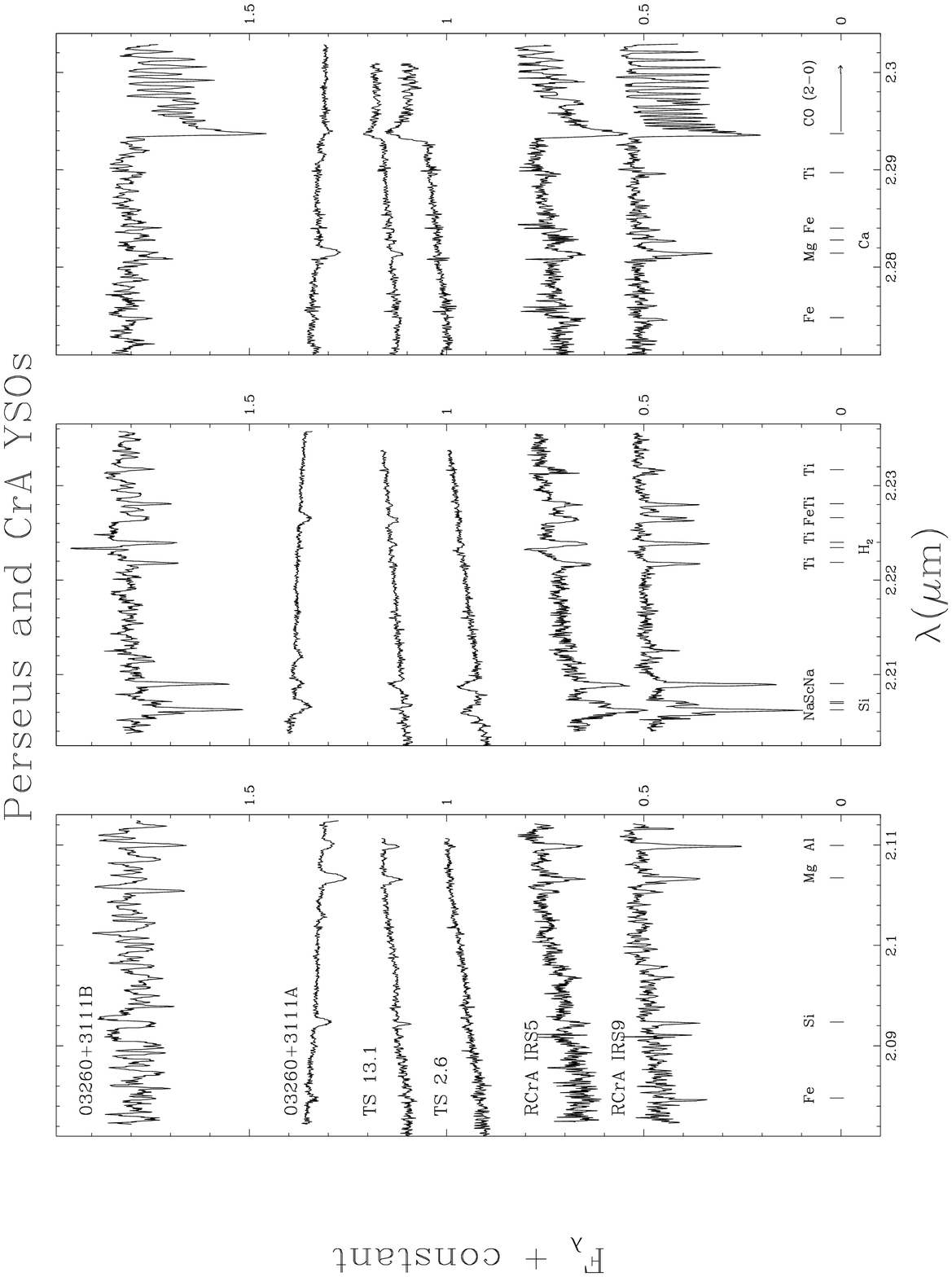}
\caption[Perseus Spectra1]
{\label{fig-all.spectra8}
High resolution (R $\simeq$ 18,000) spectra of protostars in Perseus
and Corona Australis are shown in three echelle orders, which contain
the photospheric absorption line regions (Mg/Al, Na, and $^{12}$CO
intervals, bottom to top, respectively) that we use to derive physical
properties using spectral fits to synthetic templates.  At this
resolution, the intrinsic line widths of K-band photospheric lines are
fully resolved allowing precise fits to their line shapes.}

\end{figure}

\begin{figure}
\figurenum{5}
\plotone{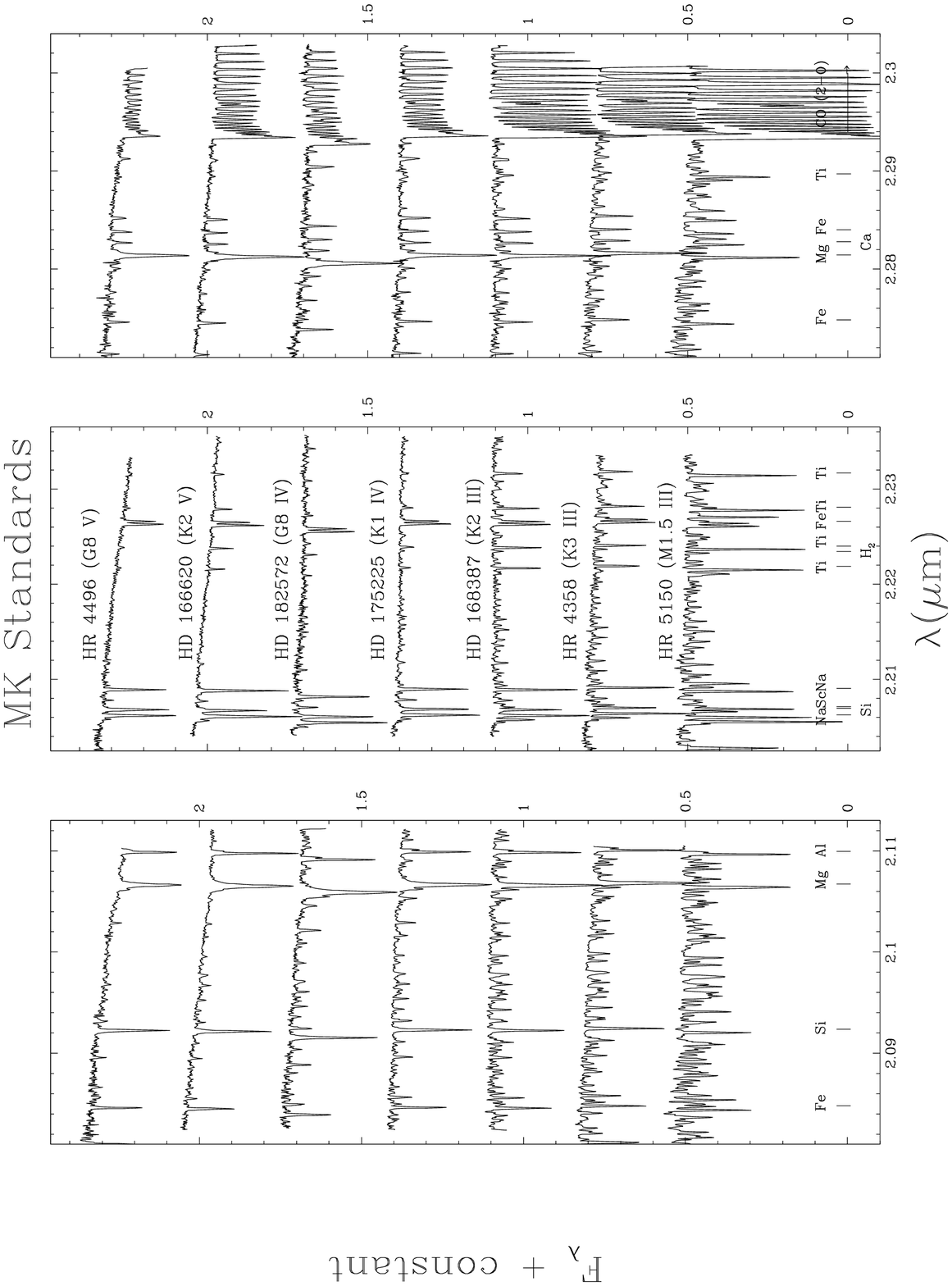}
\caption[Standards Spectra1]
{\label{fig-stands.spectra1}
Our NIRSPEC spectra (R $\simeq$ 18,000) of late-type MK standards
display visible changes in the strength of K-band absorption lines
with changes in luminosity class.  The $^{12}$CO overtone bandhead at
2.2935 $\mu$m (top panel) and Mg line triplet at 2.1066 $\mu$m (bottom
panel) show a general increase in line strength going to earlier
luminosity class (i.e. decreasing surface gravity).  Conversely, Na
lines at 2.2062 \& 2.2090 $\mu$m (middle panel) become narrower as
gravity decreases.  We make use of the Na and Mg lines together to
diagnose surface gravity in YSOs, since they display opposite
trends in their line strengths with changes in $\log g$, and are less
likely to be contaminated by non-photospheric emission lines compared
to $^{12}$CO absorption lines (see \S~\ref{sec-diaglines}).}
\end{figure}

\begin{figure}
\figurenum{6}
\plotone{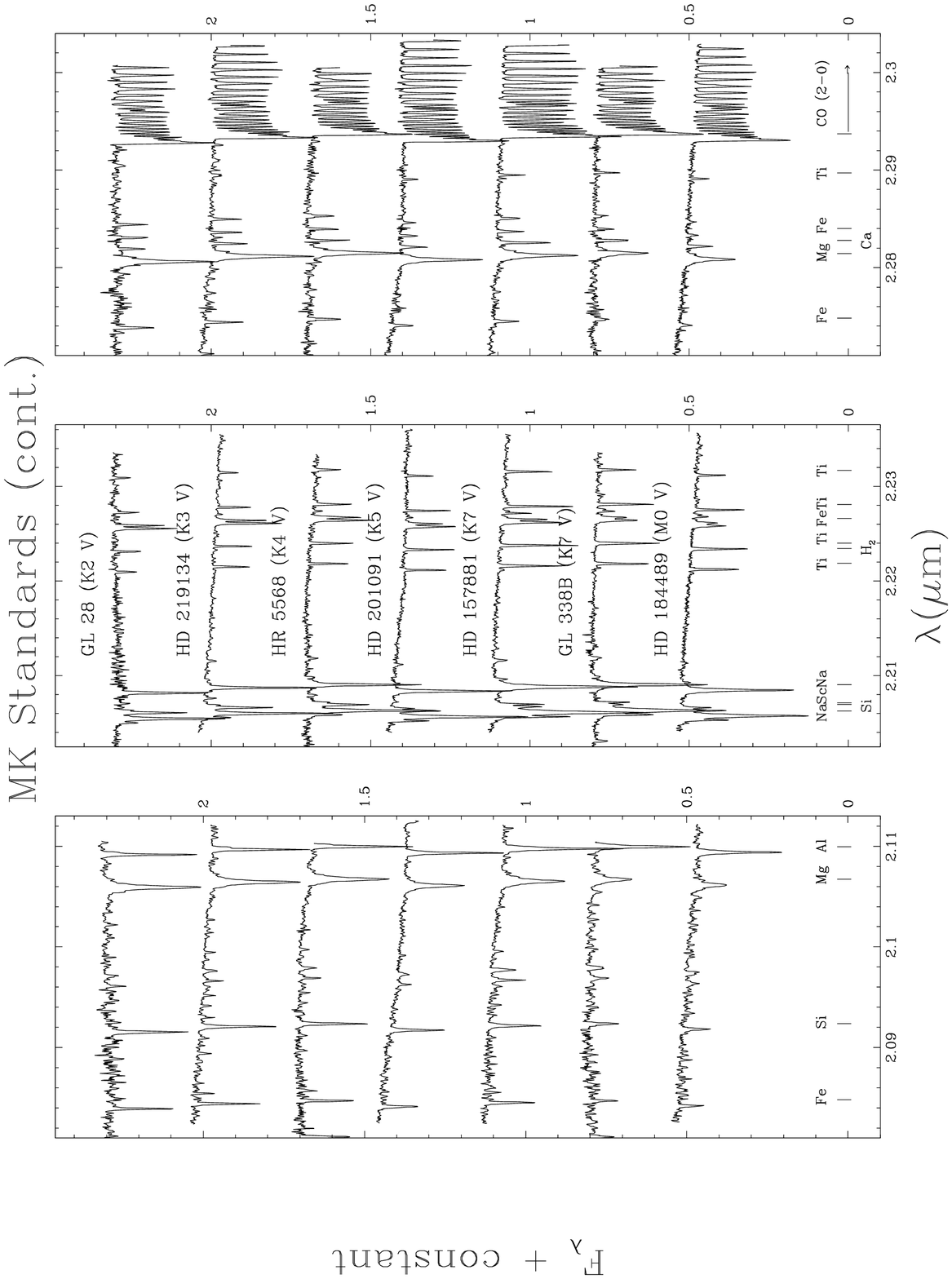}
\caption[Standards Spectra2]
{\label{fig-stands.spectra2}
Spectral trends in the absorption lines we have identified to diagnose
temperature are visible in this sequence of dwarf spectra (K2 - M0) in
MK standard stars.  In this range of fixed luminosity class, the Mg
line strength (bottom panel sequence) decreases with later
spectral-type (toward cooler effective temperatures), while the Na
lines at 2.2062 \& 2.2090 $\mu$m (middle panel) increase in strength.
Note that CO lines (top panel) change very little across this spectral
range (T$_{\rm eff}$: 5200K -- 3800K).  }

\end{figure}

\begin{figure}
\figurenum{7}
\plotone{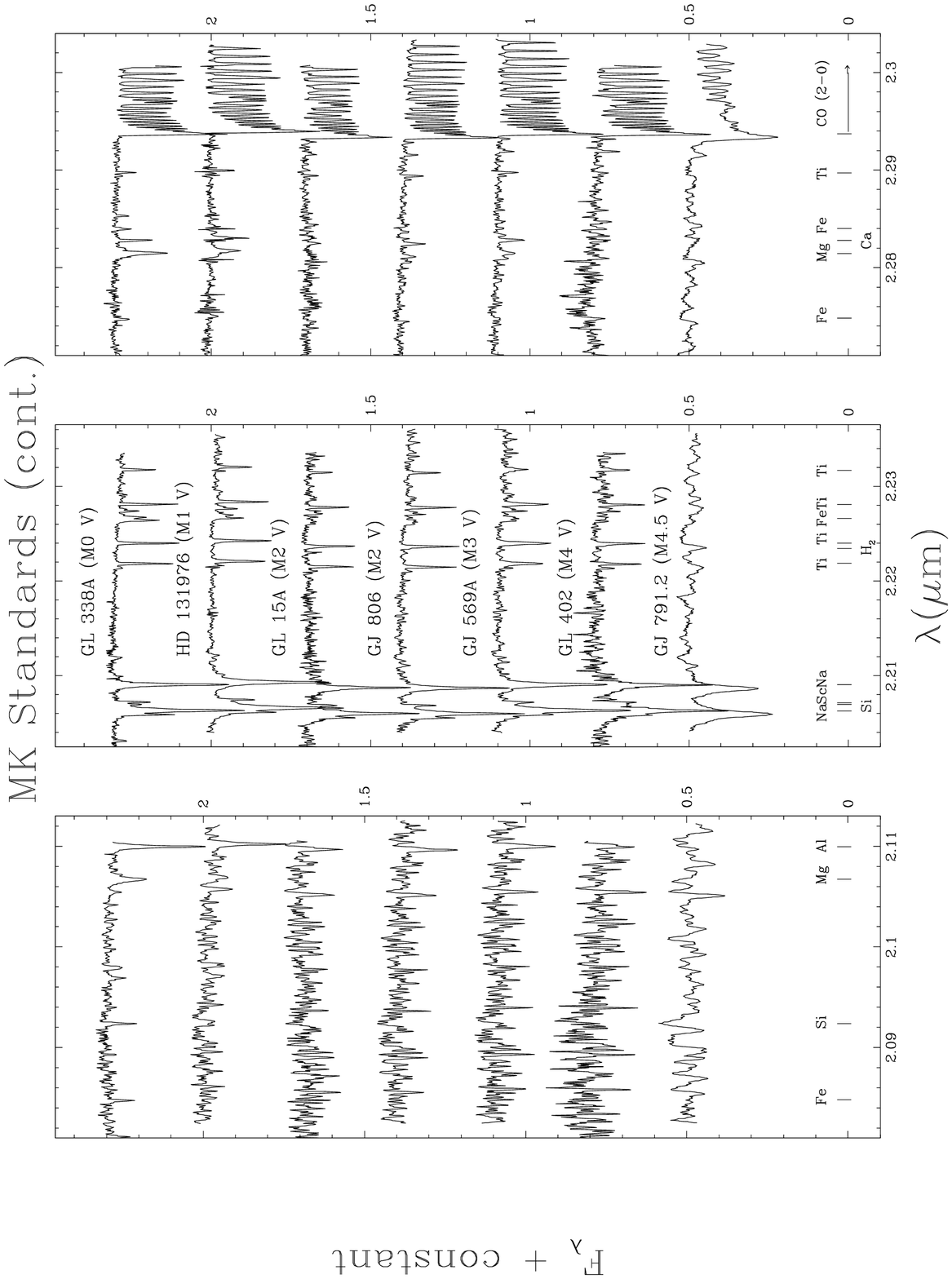}
\caption[Standards Spectra3]
{\label{fig-stands.spectra3}
This spectral sequence from early- to mid-M spectral class (M0 --
M4.5) in MK standards illustrates the inverse temperature sensitivity
of Mg (2.1066 $\mu$m) and Al (2.1099 $\mu$m) lines (bottom panel)
compared to the Na lines (2.2062 \& 2.2090 $\mu$m, middle panel).  As in
Fig. \ref{fig-stands.spectra2}, the Na lines continue to deepen and
get broader with decreasing temperature.  While the Mg line weakens
and disappears after M1, followed by the Al line which is no longer
visible after M4.  In contrast, the CO overtone band at 2.2935 $\mu$m
(top panel) shows little change in strength.  We therefore use the Na,
Mg, and Al lines together as a strong diagnostic for temperature in
late-type stellar spectra (\S~\ref{sec-diaglines}).}

\end{figure}

\begin{figure}
\figurenum{8}
\plotone{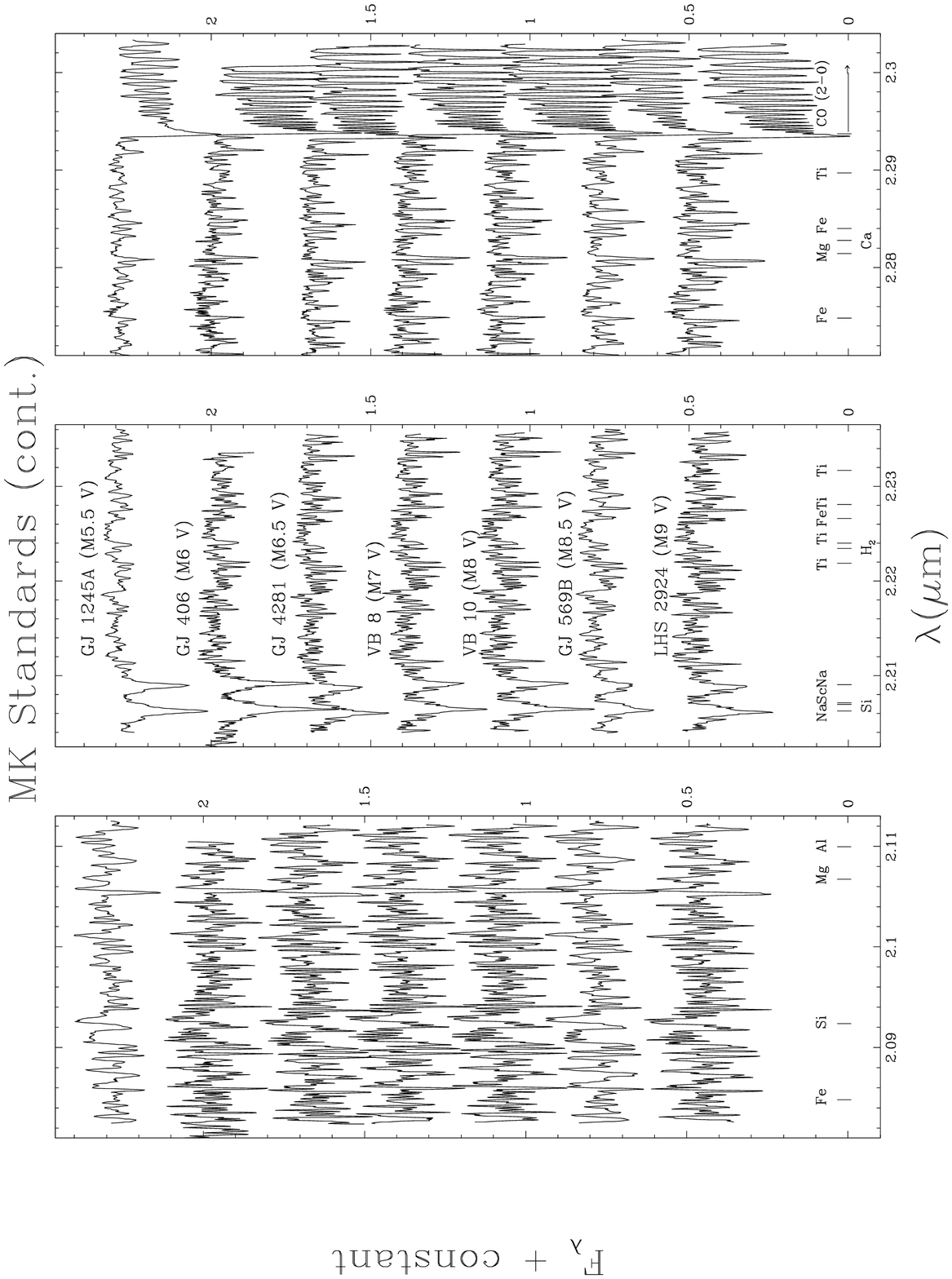}
\caption[Standards Spectra4]
{\label{fig-stands.spectra4}
As effective temperatures transition from the low mass stellar regime
into the brown dwarf regime, Mg and Al lines (2.1066 $\mu$m \& 2.1099
$\mu$m) in the bottom panel are now both absent in this decreasing
spectral sequence.  The Na lines (middle panel) no longer increase in
strength as temperatures fall below the hydrogen burning limit
(T$_{\rm eff} <$ 2900K, $<$ M6.5).  At these substellar
temperatures, Mg, Al, and Na lines are no longer useful as temperature
indicators in late-type spectra.}

\end{figure}

In order to view the true continuum shape within the three K-band
orders where we analyze photospheric lines, we multiply the spectra of
our YSOs and MK standards by a 10,000 K blackbody following the
telluric division by a hot (B7 -- A2) telluric standard star.  All
these spectra are displayed in Appendix \ref{sec-spectra}
(Figs. \ref{fig-all.spectra1} -- \ref{fig-stands.spectra4}).


\section{Results and Analysis}

\subsection{Basic Spectral Properties of Sample}\label{sec-specprops}

Our high resolution spectra reveal numerous absorption and emission
lines in our Class I and flat-spectrum sample (listed in Table
\ref{tbl-1}).  In 41/52 sources, we detect photospheric absorption
lines ($\S$ \ref{sec-diaglines}) sufficient to carry out an analysis
of their stellar properties.  Additionally, ten of these sources show
HI Br~$\gamma$ absorption, superposed on a Br~$\gamma$ emission peak
in 6/10 cases (see $\S$ \ref{sec-brgemis}).  The existence of infrared
emission lines indicative of accretion, in-fall, and outflows may
distinguish between Class I and Class II YSO populations
\citep{greene1996}.  Our flat-spectrum / Class I study shows the
presence of such lines in some of our sample: 34/52 have HI
Br~$\gamma$, 8/52 show $^{12}$CO overtone emission, and 23/52 have
H$_2$ 1-0 S(0) emission.  Table \ref{tbl-2} lists these statistics
grouped by individual star forming region.  In the 11 sources for
which we do not detect photospheric lines sufficient for our following
analysis, 10 show Br~$\gamma$ emission, while the remaining one
(CRBR~85) has marginal CO absorption evident in its lower
signal-to-noise spectrum.

In this Section 3, we describe the method by which we compare spectral
synthesis models to observed protostellar spectra and produce the best
multi-line fit to numerous K-band lines that diagnose fundamental
stellar or circumstellar properties. In particular, the resolved line
shapes, depths, and positions permit us to measure the effective
temperature (T$_{\rm eff}$), surface gravity ($\log~g$), rotational
broadening ($v\sin~i$), and radial velocity (v$_{\rm rad}$) of an
embedded YSO, as well as the amount of continuum veiling (r$_{\rm
K}$~$\equiv$~F$_{\rm Kex}$/F$_{\rm K*}$, the ratio of non stellar
excess over the stellar flux at K) arising in its circumstellar disk.
This method is based on a routine originally presented in DJ03, but
has been substantially expanded to include simultaneous fitting of
multiple wavelength regions and is thus discussed here in detail.
$\S$ \ref{sec-diaglines} describes the photospheric lines utilized by
this spectral fitting routine and the stellar parameters to which each
line is most sensitive.  $\S$ \ref{sec-synthgrid} explains the
creation of a grid of synthetic spectra for various values of T$_{\rm
eff}$ and $\log~g$, while $\S$ \ref{sec-fittingroutine} reviews the
routine which extends the spectral grid to cover a range of $v\sin~i$,
r$_{\rm K}$, and v$_{\rm rad}$, as well as selecting the set of
parameters which minimizes the overall residuals between the model
grid and the observed protostellar spectrum to find the best-fit
stellar parameters. In $\S$ \ref{sec-qualityfits} we explore the
sensitivity of the model results to both random and systemic errors,
while in $\S$ \ref{sec-standsfits} we verify the accuracy of the
stellar parameters determined by the fitting routine for spectra of
well-studied MK standards.  $\S$ \ref{sec-radvels} corrects the
routine for a slight ($< 2 \rm {~km~s}^{-1}$) bias in the detected
radial velocity.  Finally, in $\S$ \ref{sec-photlums} we describe the
way we derive stellar luminosities from photometry, pointing out
issues and uncertainties associated with using this traditional
approach for placing YSOs on the H--R diagram.

\subsection{Diagnostic Lines}\label{sec-diaglines}

We focus our analysis on key diagnostic absorption lines whose
sensitivity to stellar parameters we have identified and tested with
late-type spectra of MK dwarfs and giants.  The cross dispersed
wavelength coverage afforded by NIRSPEC enables the detection of
photospheric lines across multiple echelle orders.  We select strong
absorption lines of neutral metals and the molecular overtone band of
(2-0) $^{12}$CO, present in three K-band echelle orders, to diagnose
the physical state of our embedded protostars, by measuring effective
temperature (T$_{\rm eff}$), surface gravity ($\log~g$), projected
rotation ($v\sin~i$), continuum veiling (r$_{\rm K}$), and radial
velocity (v$_{\rm LSR}$).

At 2.21 $\mu$m (order 34), we utilize neutral atomic lines of Na, Si,
and Sc (hereafter, the Na interval), proven diagnostics of T$_{\rm
eff}$ in MK standard stars (DJ03).  At 2.11 $\mu$m (order 36), lines
of neutral Mg and Al (hereafter, the Mg/Al interval) are also present
in K and M spectral-type stars, and are sensitive to changes in
T$_{\rm eff}$ and $\log~g$.  The first overtone of the $^{12}$CO
bandhead at 2.29 $\mu$m (order 33, hereafter the CO interval) is also
prominent in the photospheres of cool, late-type stars.  Changes in
T$_{\rm eff}$ and $\log~g$ affect the depth of the bandhead and R
branch lines, without broadening the feature (i.e no pressure
broadening).  Furthermore, CO is not affected by Zeeman broadening
seen in K-band atomic lines of T-Tauri stars with strong magnetic
fields \citep{johnskrull1999,johnskrull2001}.  Therefore, our most
reliable measurement of $v\sin~i$ in our protostars comes from
spectral fits to the CO interval.

At cool effective temperatures (T$_{\rm eff}$: 3200K -- 4500K) and
sub-dwarf surface gravities ($3.5 \le \log~g \le 4.5$), the Na lines
in the Na interval grow and broaden as T$_{\rm eff}$ decreases and
$\log~g$ \textit{increases} (see Figs. \ref{fig-stands.spectra1} \&
\ref{fig-stands.spectra2} in Appendix \ref{sec-spectra}).  Conversely,
the Mg lines (a close triplet) in the Mg/Al interval are degenerate
between temperature and gravity, but in the opposite sense as are the
Na lines (i.e. a decrease in T$_{\rm eff}$ trades off with a
\textit{decrease} in gravity).  Earlier studies have identified the
strength of the CO overtone band at 2.2935 $\mu$m as being diagnostic
of luminosity and surface gravity \citep{baldwin1973, kleinmann1986,
lancon1992,ramirez1997} in stars that have already evolved to the
main-sequence and beyond, and in Class II PMS stars (DJ03) that are
less likely to have contamination by CO emission arising from disks or
in outflow winds.  For our Class I and flat-spectrum YSOs, we avoid
using fits to the CO interval to constrain temperature and gravity,
since the strength of the bandhead absorption may be altered by
non-stellar components.

\subsection{A Grid of Synthetic Spectra}\label{sec-synthgrid}

We develop a multi-parameter grid of synthetic spectra in which we
search to find the best simultaneous fit to the wavelength intervals
described in $\S$~\ref{sec-diaglines}, extending the spectral fitting
techniques pioneered by DJ03 to include these three distinct
wavelength regions.  The spectral synthesis code MOOG
\citep{sneden1973} is used to generate a high resolution (R~=~
120,000) K-band spectrum of each wavelength interval based on LTE
radiative transfer through a NEXTGEN \citep{hauschildt1999} stellar
atmosphere profile (see DJ03).  Each MOOG synthesis generates a model
spectrum specific to a particular T$_{\rm eff}$, $\log~g$, and
metallicity.  We restrict our spectral search to objects with solar
abundances \citep{padgett1996}, though, in principle, spectral fits
using metallicity as another free search parameter have potential for
future work when more diagnostic lines are present.

Physical constants (i.e. oscillator strengths and damping
coefficients) are needed in order to model the correct depth and width
for each absorption line transition in the synthetic spectra.  For our
spectral syntheses in the Na and CO intervals, we use atomic and
molecular line constants published by \citet{kurucz1994} and
\citet{goorvitch1994}, respectively.  However, no published oscillator
strengths or damping coefficients were available in the literature or
on-line (i.e. VALD or NIST) for the Mg line transitions
(3s4f$^3$F$^0_{2,3,4}$--3s7g$^1$G$^3_{3,4,5}$) at 2.1061 $\mu$m in the
Mg/Al interval.  Therefore, we determined these values empirically by
fine tuning the models to best fit our observed spectra of several MK
standard stars for which T$_{\rm eff}$, $\log~g$, and [Fe/H] was
known.

We use solar micro-turbulence values (1 km~s$^{-1}$), appropriate for
our models of late-type stars with dwarf and sub-dwarf gravities
(DJ03).  Damping of the Na lines, which are noticeably pressure
broadened in the Na interval in dwarf standards, has been tuned to
best match an observed solar spectrum (DJ03).

\subsection{Spectral Fitting Routine}\label{sec-fittingroutine}

Once the basic synthesis grid has been created, a spectral search
routine (DJ03) steps through each model spectrum, modifying it by
incrementally adding rotational broadening ($v\sin~i$), continuum
veiling (r$_{\rm K}$), and fractional pixel wavelength shifts ($\Delta
v_{\rm rad}$). Broadening by stellar rotation is modeled by convolving
our model spectra with a rotational broadening profile and an assumed
limb-darkening coefficient of 0.6 \citep{gray1992}.  We simulate the
amount of 2 $\mu$m continuum veiling (r$_{\rm K}$) by adding
successive amounts of non-photospheric flux, assumed to be spectrally
flat and featureless.  A coarse value for the topocentric radial
velocity is estimated by eye for each observed spectrum.  Then, 0.8
km~s$^{-1}$ shifts ($\sim$1/5 of a pixel) are each tested over a
$\pm$5 km~s$^{-1}$ range to find the best radial velocity fit to each
protostellar spectrum.

\begin{figure}
\figurenum{9}
\plotone{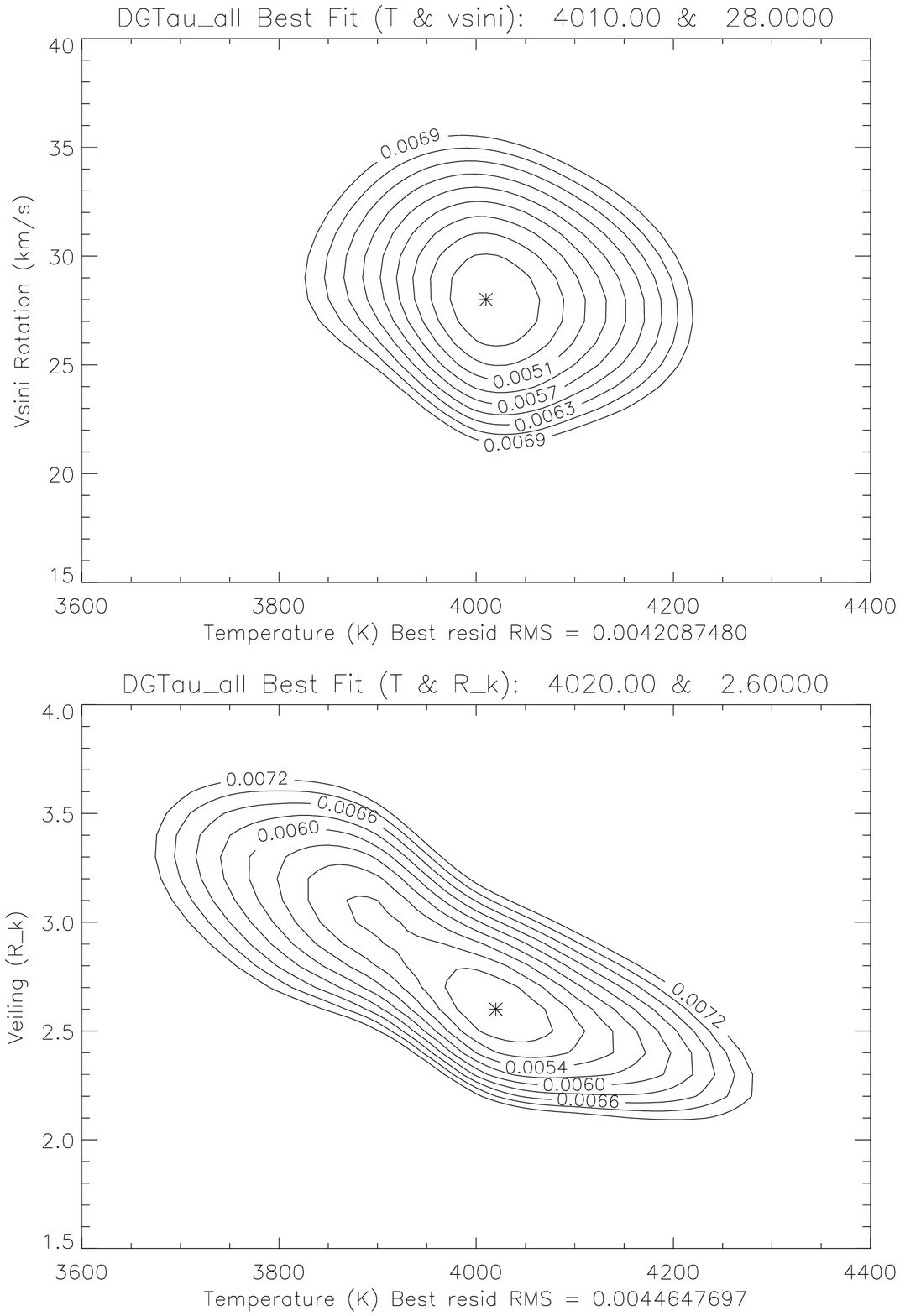}
\caption[HRD logg]
{\label{fig-rmscontour.teff} 
Typical uncertainties are displayed in the derived values of T$_{\rm
eff}$, $v\sin~i$, and r$_{\rm K}$ with spectral fits to DG Tau.  The
shape and density of the error space contour planes corresponding to
the model fits of these three physical parameters illustrate how
increased veiling and cooler effective temperatures are degenerate
with one another by the elongated 3~$\sigma$ contour intervals (lower
panel).  In contrast, the upper panel shows a symmetric minimum in the
best-fit value for rotation and effective temperature, indicating that
these two parameters show more distinguishable behavior in their line
shapes with changes in these values (see table \ref{tbl-3}).  }
\end{figure}

The search routine steps through a selected region of parameter space,
modifying the spectral synthesis templates of all three wavelength
intervals by modeling each unique combination of the five search
parameters within the following range of values: 3000K -- 6000K
(T$_{\rm eff}$), 3.5 -- 5.0 cm s$^{-2}$ ($\log~g$), 0 -- 80
km~s$^{-1}$ ($v\sin~i$), 0 -- 6.0 (veiling, r$_{\rm K}$), and $\pm$5
km~s$^{-1}$ (v$_{\rm rad}$).  Each unique model spectrum is compared
to the observed YSO spectrum.  Wavelength sub-intervals in the
observed spectrum are selected by eye to isolate only those
photospheric lines which will be compared to their model counterparts.
The quality of the fit between the observed protostellar spectrum and
each spectral synthesis model is then quantified from the
aforementioned selected wavelength sub-intervals as the root mean
square (RMS) of the residuals in the data-to-model comparison (DJ03).
The overall best-fit model is automatically selected by the search
routine as the one with the lowest RMS value over the entire search
grid.  Fig. \ref{fig-rmscontour.teff} illustrates the RMS error space
in the veiling/effective temperature plane for the model fits to
DG~Tau.  The elongation in the error space stretching from upper left
to lower right in the lower panel shows that there is still a bias for
lines with higher amounts of veiling to trade off with lines of cooler
temperature stars.

The range of the search grid in T$_{\rm eff}$ and $\log~g$ is
constrained by the availability of the NEXTGEN models
\citep{hauschildt1999} which we use to generate the spectral synthesis
templates.  RMS values corresponding to each of the fits in the three
wavelength intervals are combined in the following way after
validating this technique with the fits to MK standards of known
T$_{\rm eff}$, $\log~g$, and $v\sin~i$ ($\S$ \ref{sec-standsfits}).
(1) We use the Na and Mg/Al wavelength intervals together as combined
diagnostics of temperature, gravity, and veiling in an iterative
process.  Initially, gravity is held fixed, while the best fit for
temperature and veiling is found.  Then, temperature is fixed at the
previous best-fit value while gravity and veiling were varied to find
the best fit.  This process is iterated until a best fit for all three
parameters is found.  By fitting both Na and Mg/Al wavelength
intervals simultaneously we usually converge to a unique value for
effective temperature, gravity and veiling where the combined RMS
value from both Na and Mg/Al intervals is lower and the error space
more sharply defined than when either interval is used alone.
Therefore, we use both intervals weighted together, equally, as strong
diagnostics for T$_{\rm eff}$ and $\log~g$, as well as r$_{\rm K}$.
(2) The broadening of the bandhead in the CO interval is determined to
be the most sensitive indicator of projected $v\sin~i$ rotation of all
our diagnostic lines.  Therefore $v\sin~i$ is solely determined by the
best fits to the CO interval by first imposing the best T$_{\rm eff}$
and $\log~g$ from the Na and Mg/Al intervals, while leaving the amount
of veiling unconstrained, in favor of finding the best rotationally
broadened spectral fit.  This amount of broadening is likewise imposed
on the fits in the other two intervals.

Radial velocities for each object are found as the mean of the
velocity shifts between the observed and the best-fit synthetic
spectra for each of three spectral regions independently.
Occasionally one of the three orders had a widely divergent derived
velocity shift; in these cases the discrepant point is discarded from
the mean. The derived velocities are adjusted using the IRAF package
RVCORRECT to account for motions due to the Earth's solar and galactic
orbits, allowing a calculation of V$_{\rm Helio}$ and V$_{\rm LSR}$
for each object.

\subsection{Quality of the Model Fits}\label{sec-qualityfits}

\begin{figure}
\figurenum{10}
\plotone{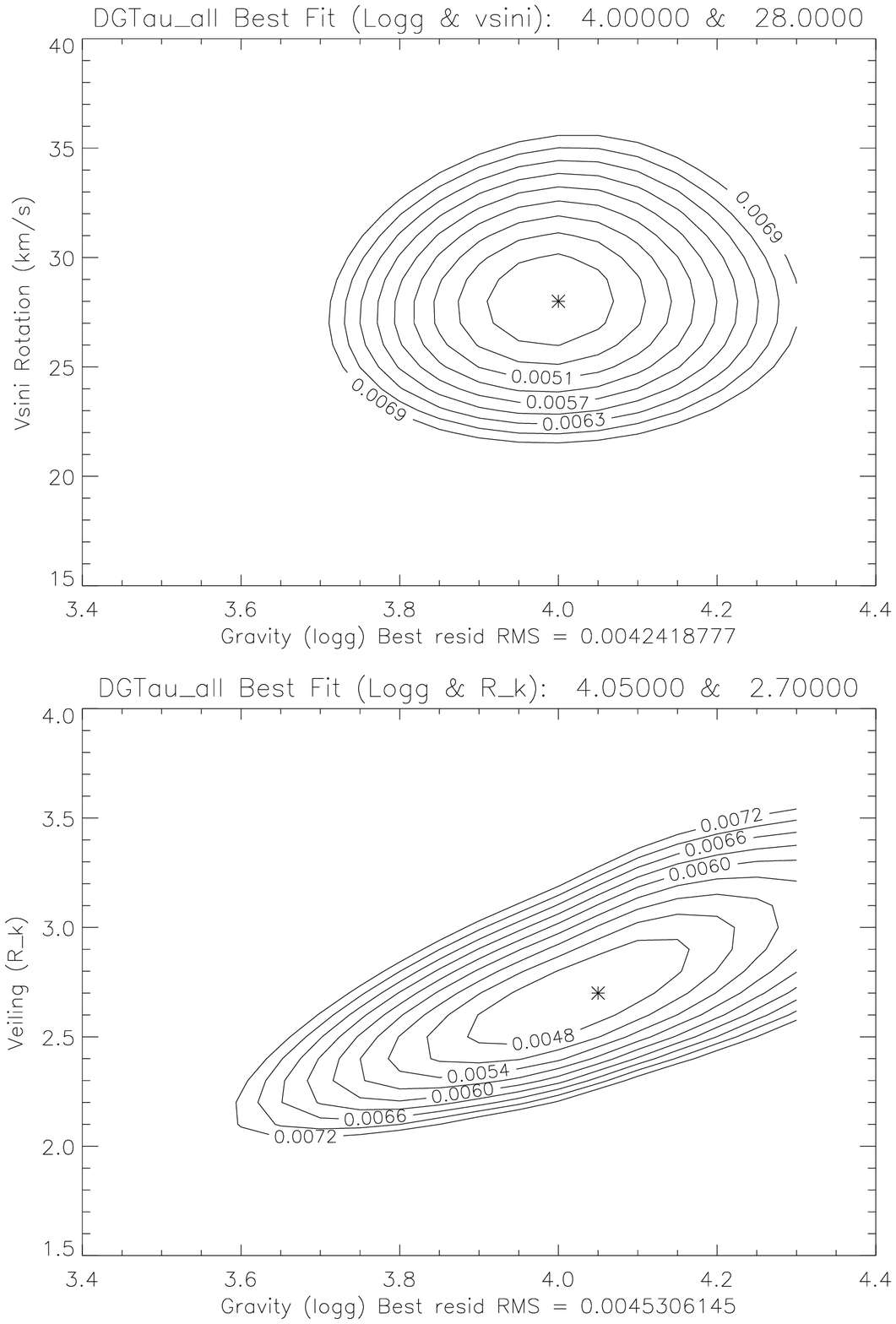}
\caption[HRD logg]
{\label{fig-rmscontour.logg} 
Typical uncertainties in surface gravity, $v\sin~i$, and r$_{\rm K}$
are shown here in the error space of spectral fits to DG Tau.  Surface
gravity and veiling are degenerate with one another, as indicated by
the elongated 3 $\sigma$ RMS error contour intervals (lower panel),
but in the opposite sense as with T$_{\rm eff}$ vs.  veiling (fig.
\ref{fig-rmscontour.teff}).  By fitting for gravity and T$_{\rm eff}$
simultaneously then errors in veiling are significantly reduced.  (see
table \ref{tbl-3})}
\end{figure}

Random errors present in the RMS best-fit value of the spectral fits
are due to the finite signal-to-noise present in our data.  In order
to calibrate a 3~$\sigma$ change in the RMS parameter search space, a
test model spectrum was constructed using a particular synthesis
model, representative of a typical YSO fit in our sample: (T$_{\rm
eff}$ = 3600K, $\log~g$ = 4.0, $v\sin~i$ = 40, and r$_{\rm K}$ =
3.0). Random Gaussian noise was added to the test spectrum to achieve
S/N~=~200 pixel$^{-1}$ before it was then run through the standard
fitting routine.  This was done many times with random noise seeds.  
A 1~$\sigma$ variation in the best-fit RMS value over many
trial fits to the same spectrum with noise was $\Delta$RMS =
10$^{-4}$.  Using this number, we calibrate the errors due to noise in
each of the physical parameters using the RMS error contour plots
(output by the spectral fitting routine for each fit).  A change in
the RMS from the lowest value (i.e. the best fit) to a value that is
bigger by 3 $\times$ 10$^{-4}$, then corresponds to the 3~$\sigma$
random error along that new error contour line.  For example,
Fig. \ref{fig-rmscontour.logg} shows the error space contours in the
$\log~g$ vs r$_{\rm K}$ and $\log~g$ vs $v\sin~i$ planes corresponding
to the best fits of these quantities in DG~Tau.  The asterisks mark
the minima in the error space (RMS = 4.2 $\times$ 10$^{-3}$ and 4.5
$\times$ 10$^{-3}$) and each contour interval corresponds to an
\textit{increase} of 3 $\times$ 10$^{-4}$ (3~$\sigma$ contours),
resulting in a calibrated change for each parameter.  Therefore, the
error space in each of four parameter fit planes (T$_{\rm eff}$ vs
$v\sin~i$), (T$_{\rm eff}$ vs r$_{\rm K}$), ($\log~g$ vs $v\sin~i$),
($\log~g$ vs r$_{\rm K}$) is characterized by changes of 10$^{-4}$
(1~$\sigma$) from the best-fit RMS value.  Random errors in the four
parameters are generally smaller than systematic errors.

The best model fit to each observed protostellar spectrum is
associated with an initial seeded $\log~g$ value, where the initial seed
value for $\log~g$ is assumed and held fixed while the spectral
fitting routine then finds the best-fit T$_{\rm eff}$.  Following
this, the routine then holds the new best-fit T$_{\rm eff}$ fixed and
searches for the corresponding best fit for $\log~g$, and so forth.
Thus, the routine finds an RMS minimum within the 5 parameter search
space, iterating between the T$_{\rm eff}$ and $\log~g$ search planes.
Sometimes there is more than one local minima in the solution and
different initial seed values for $\log~g$ fit to different best-fit solutions.
In this case, differences in the different RMS minima are compared and
the overall best solution is taken as the global minimum of all
different seeded best fits.  Uncertainties (1 $\sigma$) in the
physical parameters (T$_{\rm eff}$, $\log g$, r$_{\rm K}$, and $v~\sin
i$) derived from the spectral fits are computed by measuring the
changes in each quantity that produced a change of $10^{-4}$ in the
RMS fit residual, about the global minimum RMS fit.  Systematic errors
in the best-fit parameters are then gauged by changes in best-fit
solutions that are below our ability to discern real changes in the
fits appropriate to the signal-to-noise of our spectra.
 
Changes in the equivalent widths of lines are evident in the spectra of
young stars with differing T$_{\rm eff}$, $\log~g$, and r$_{\rm K}$.
At high resolution the changes to the lines whose shapes are fully
resolved can distinguish between different physical diagnostics
(i.e. line damping and large $v\sin~i$ rotation), but degeneracies in
the fits to spectral models in T$_{\rm eff}$, $\log~g$, and r$_{\rm
K}$ persist even at high spectral resolution, especially in
unsaturated lines.  The elongation in the error space in figures
\ref{fig-rmscontour.teff} and \ref{fig-rmscontour.logg} (bottom
panels) show how T$_{\rm eff}$ and $\log~g$ are both somewhat
degenerate with veiling, though in opposite senses.  Also, surface
gravity and effective temperature are together degenerate (DJ03) in
the sense that lower effective temperatures fit to lower $\log~g$
spectra (see further discussion in $\S$ \ref{sec-standsfits}).

The degeneracy is less pronounced when simultaneous fits to more
spectral lines can be performed, as has been done here with the
NIRSPEC echelle spectra.  In particular, the Na fits and Mg/Al fits
together constrain gravity and temperature better than using only one
interval, despite systematic errors that can dominate the accuracies
of the best-fit model.

\subsection{Spectral Fits}\label{sec-standsfits}

\begin{figure}
\figurenum{11}
\plotone{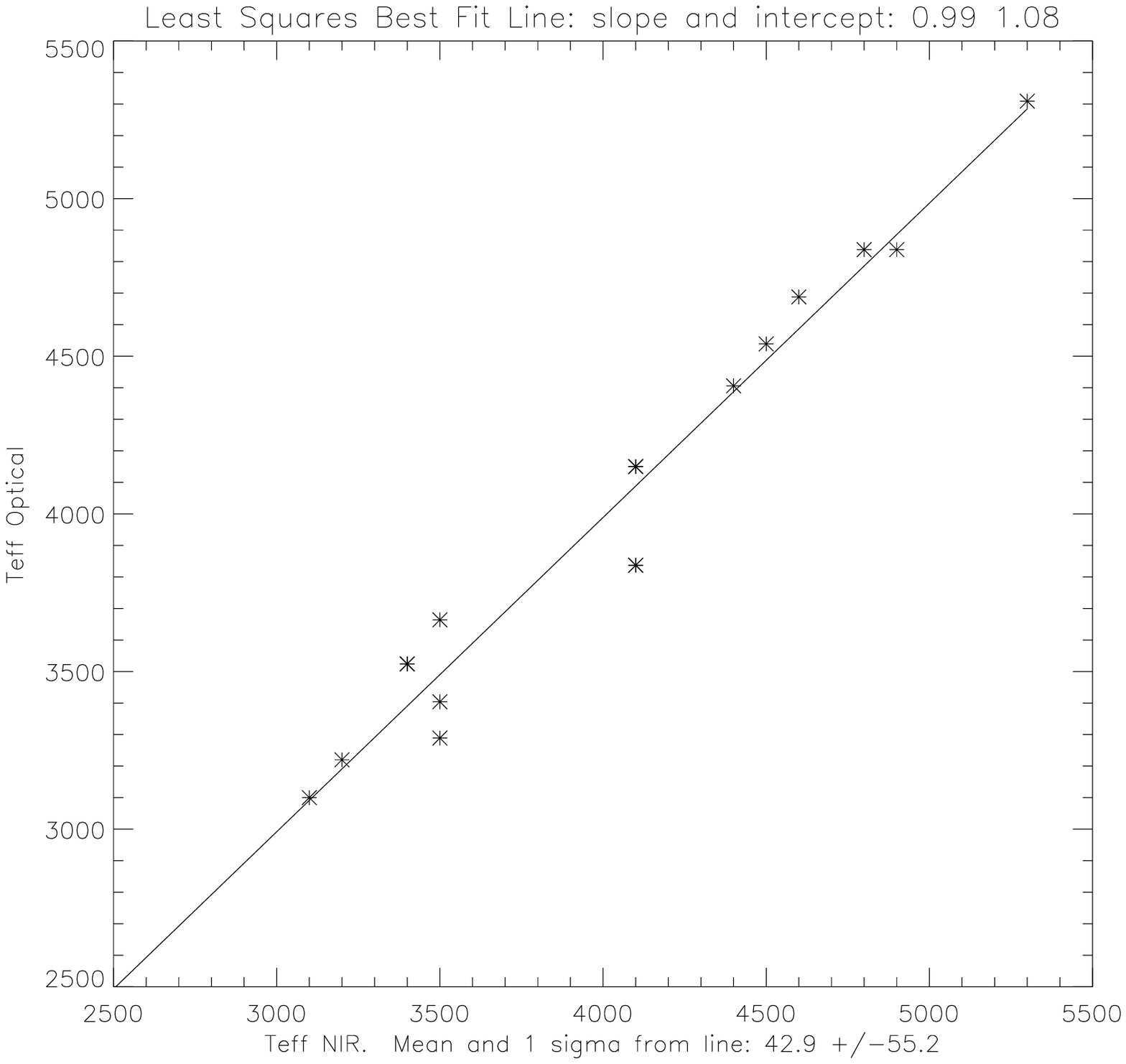}
\caption[teff Fits to Stands]
{\label{fig-teff.fitstands}
Derived effective temperatures of MK standards
based on fits to spectral synthesis models using NIR photospheric
lines (horizontal axis) are plotted against their optical temperatures
as derived from their published spectral-types using
a SpT / T$_{\rm eff}$ relation given by \citet{dejager1987}.  The linear least squares
fit (solid line) shows a very close fit (slope = 0.99,
and y-intercept = 1.08K) to that of a perfect fit line,
(T$_{\rm eff~NIR} \equiv$ T$_{\rm eff~optical}$).
The sources show a 1~$\sigma$ scatter of 57K consistent about the
best-fit line.  A small amount of veiling (mean r$_{\rm K}$ = 0.2) was needed
in our best-fit spectral models to fit these standards, which
we have corrected for when we fit the YSOs in our sample. 
}

\end{figure}

To validate the quality of the spectral synthesis models and our
ability to accurately fit them to actual data, we apply our fitting
routine ($\S$ \ref{sec-fittingroutine}) to our NIRSPEC observations of
MK standards.  Using our standard fitting procedure to the synthesis
model grid, we determine the best-fit values for T$_{\rm eff}$
spanning G8 -- M5.5 spectral-types (5300K to 3100K) in our standard
stars.  We find a very good agreement (correlation coefficient of
0.99) between our near-IR temperatures and those inferred from the
optical spectral-types (Fig. \ref{fig-teff.fitstands}) using the
relation quoted by \citet{dejager1987}.  The surface gravities we
derive in the standards are consistent with those of late-type dwarfs
($\log~g$ = $\sim$4.4 -- 5.0) with some scatter due to the
uncertainties in the conversion from the standards' spectral-types to
surface gravities.  All but 2 of our late-type standards have
$v\sin~i$ rotational velocities well below our resolution limit (R =
18,000, 17 km~s$^{-1}$). The average $v\sin~i$ value derived for these
slow rotating standards is 10.8 km~s$^{-1}$, which we adopt as a
measure of the instrumental broadening through our 4 pixel wide slit.
We remove this value in quadrature from the $v\sin~i$ values we
measure in two standards (Gliese 791.2 \& 1245A) whose velocity
broadened lines we resolve and compare with independent derived
optical measurements of 32 and 22.5 km~s$^{-1}$, respectively
\citep{mohanty2003}.  Our near-IR derived $v\sin~i$ values are in
excellent agreement with the optical study, where we measure, 34.3 and
22.5 km~s$^{-1}$, respectively.

We expect our MK standard stars that are in the field to be completely
unveiled.  However, when we allow veiling to be fit as a free
parameter along with T$_{\rm eff}$, $\log~g$, and $v\sin~i$, we find
the best-fit value has an average veiling of r$_{\rm K,fit}$ = 0.2,
weighted toward more veiling in later type spectral standards.  We
also see a similar effect in the CO order.  If this is an instrumental
effect caused by scattering or leaking some of each object's own
light, then the amount of this excess instrumental veiling is
$\Delta$r$_{\rm K} = 0.2 (1 + \rm r_{\rm K}$) where r$_{\rm K}$ is the
object's true veiling.  We correct the best-fit veiling values r$_{\rm
K,fit}$ of the YSOs (as derived from the Na and Mg/Al intervals) to
true intrinsic veilings (r$_{\rm K}$) with this relation and report
only these corrected values in this publication:

\begin{equation}
\label{eqn-veil.correct}
 \rm r_{\rm K}= \frac{\rm r_{\rm K,fit}-0.2}{1.2}
\end{equation}

After validating our technique with the standards, and establishing
the veiling correction needed in our NIRSPEC data, we fit 41 spectra
of Class I and flat-spectrum sources to the synthesis models following
the procedure described in $\S$ \ref{sec-fittingroutine}.  At least 1
echelle order displayed photospheric absorption lines in all of these
41 sources from which we were able to derive key physical properties
directly from the spectra.  These results are listed in Table
\ref{tbl-3} along with the 1 $\sigma$ uncertainties ($\S$
\ref{sec-qualityfits}).  We do not apply our fitting routine to 11
stars in our sample due to the absence of photospheric lines in their
spectra.

As a check to the effective temperatures, rotational broadening and
veiling values that we derive using our spectral synthesis models, we
evaluate fits to our protostellar spectra using the observed standard
spectra (that have been spun-up and veiled), where the best fit is
found qualitatively by eye.  We compare the best model fit to the
best standard fit after converting the spectral-type of each best-fit
dwarf MK standard to an effective temperature \citep{dejager1987}.
The $v\sin~i$ and r$_{\rm K}$ values both show some scatter that is
roughly symmetric about a perfect-fit line.  The T$_{\rm eff}$ values
show a systematic offset toward higher derived effective temperatures
($\Delta T_{\rm eff} \equiv T_{standards}- T_{models} = 190K$).  This
temperature difference is likely due to differences in surface
gravities between the dwarf standards ($\log~g \simeq$ 4.3 -- 4.7) and
our YSO models (mean $\log~g$ = 3.8).  In general, effective temperatures
increase by $\sim$ 200K for each increase of 0.5 in $\log~g$ for
late-type stars (DJ03), so the sign and magnitude of the temperature
offset is correct.

\begin{figure}
\figurenum{12}
\plotone{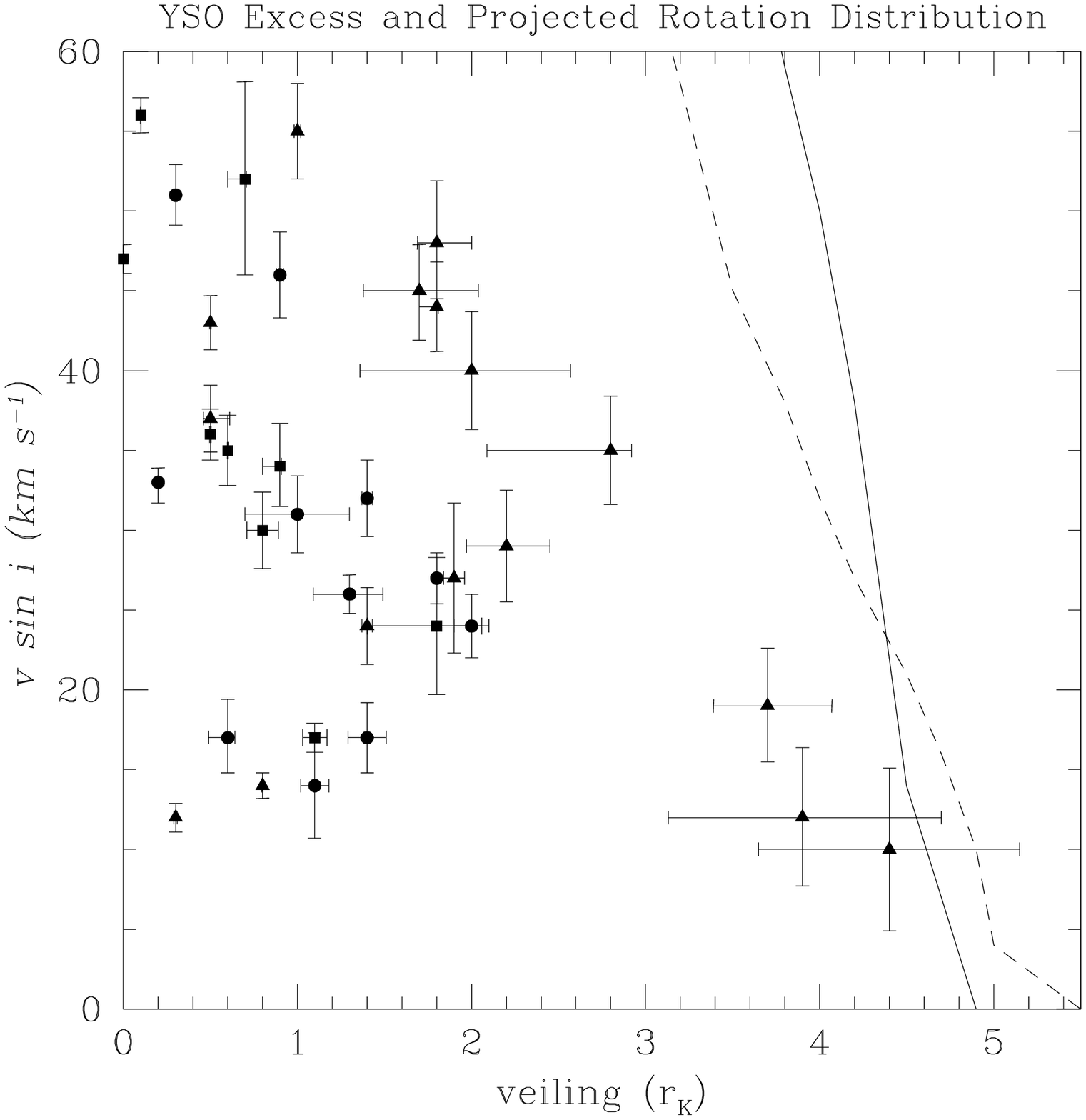}
\caption[vsini veil]
{\label{fig-vsini.veil}
Our measured $v\sin~i$ values are plotted against the continuum
veilings we derive for Class I and flat-spectrum sources in Taurus
(circles), Ophiuchus (triangles), and Serpens (squares), along with
1~$\sigma$ error bars.  The lines on the right side of the plot show
2~$\sigma$ detection limits based on the photospheric lines of two
late type MK standards (HD~131976, T$_{\rm eff}$= 3500K, solid line;
and HD~219134, T$_{\rm eff}$= 4600K, dashed line) that have been
compared to a featureless YSO in our sample (WL~6).  The spectra of
these standards was first modified by different amounts of veiling and
rotational broadening, and then Gaussian noise was added (S/N = 180,
typical in our YSO sample) to find the r$_{\rm K}$ and $v\sin~i$ pairs
that defined a 2~$\sigma$ line detection based on the RMS value (see
\S~\ref{sec-qualityfits} in the text) of each fit to WL~6.}
\end{figure}

The $v\sin~i$ rotation velocities and r$_{\rm K}$ continuum veilings
for each of the 41 objects fit in our sample (Table \ref{tbl-3}) are
shown in Figure \ref{fig-vsini.veil}. This figure also shows 2 curves
which plot our expected sensitivity to $v\sin~i$ and r$_{\rm K}$ for
T$_{\rm eff}$ = 3500K and 4600K stars with signal-to-noise
representative of our object sample (S/N = 180).  We should detect
photospheric absorption lines in objects which lie to the left of
those curves, and we are insensitive to detecting photospheric
absorption features in objects which lie to the right of these
curves. The curves and points show that we are least sensitive to
objects with high r$_{\rm K}$ and high $v\sin i$ and most sensitive to
ones with low r$_{\rm K}$ and low $v\sin~i$; high veilings and fast
rotation velocities both reduce the depths of absorption features
relative to continua.

K-band veilings and rotation velocities do not appear strongly
correlated in Figure \ref{fig-vsini.veil}. We detect objects which
span nearly the entire range of $v\sin~i$ and r$_{\rm K}$ to which we
expect to be sensitive. However, there are no objects in the region
near the line defined by the points r$_{\rm K}$ = 1, $v\sin~i$ = 0
km~s$^{-1}$ and r$_{\rm K}$ = 4, $v\sin~i$ = 40 km~s$^{-1}$. It is
unclear whether this gap is due to a physical effect or else a
characteristic of our finite object sample.

\subsection{Accuracy of Radial Velocity Measurements}\label{sec-radvels}

\begin{figure}
\figurenum{13}
\plotone{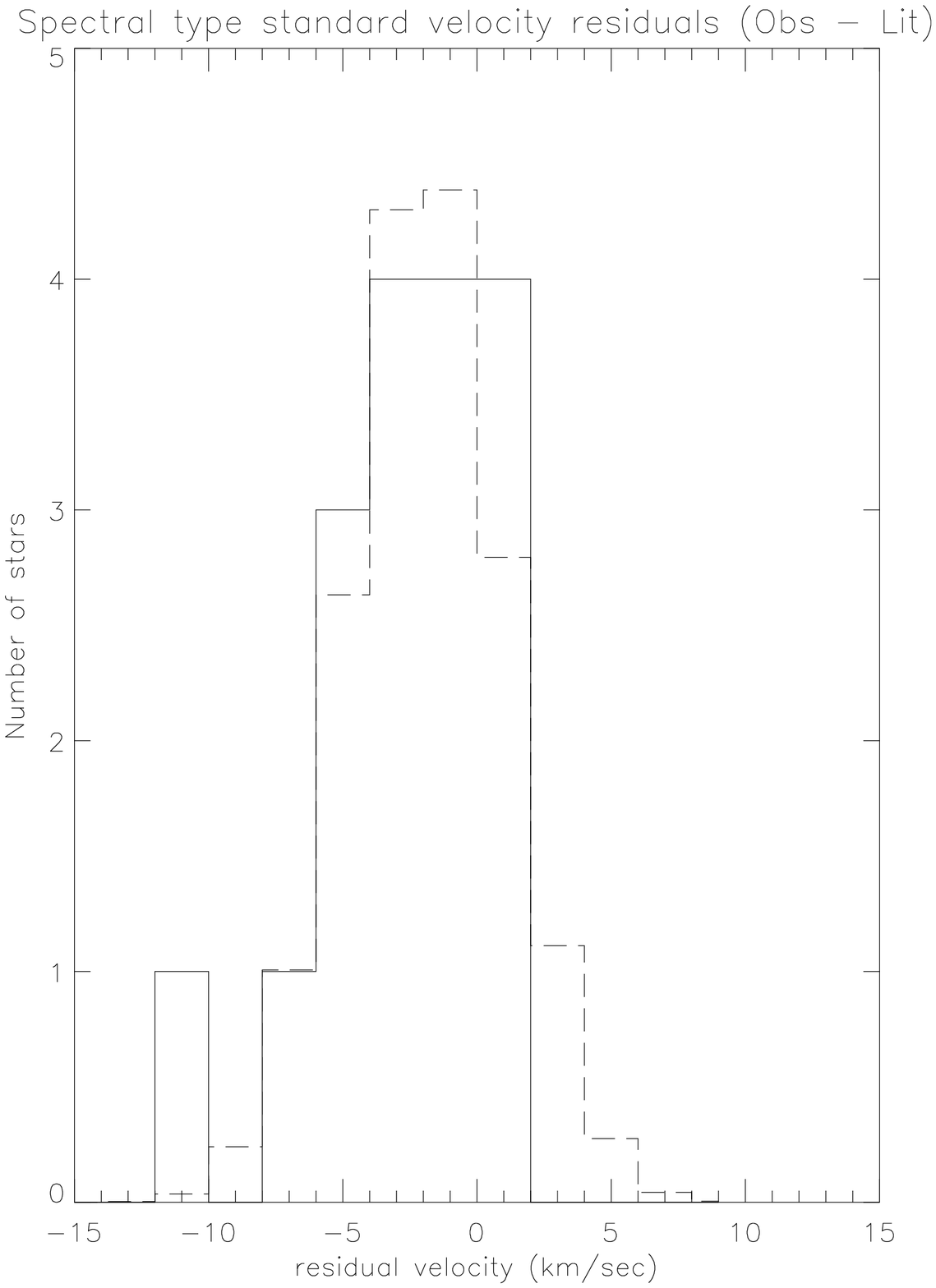}
\caption[radvel offset] {\label{fig-radvel.offset}
A histogram of radial velocity residuals (solid line) shows
differences between radial velocities measured in our standards minus
values reported in the literature for the same stars.  A Gaussian fit
(dashed line) to these residuals indicates a --1.8 km~s$^{-1}$ offset,
which we correct for in the measured radial velocities of our YSOs
(Table \ref{tbl-3}).}

\end{figure}

To determine the accuracy of our derived radial velocities, we
compare our derived heliocentric radial velocities to those in the
literature for well-observed spectral-type standards.  Figure
\ref{fig-radvel.offset} shows a histogram (solid line) of the
residuals of our derived velocities less the previously determined
value for spectral-type standards with precise (error $\le$ 2
km~s$^{-1}$) heliocentric velocities
\citep{demedeiros1999,gizis2002,nidever2002,mohanty2003}.  The best
fit Gaussian to the histogram, binned in the same manner, is shown as
the dashed line.  The mean of the Gaussian fit is --1.8 km~s$^{-1}$,
indicating the presence of a slight offset in our derived velocities.
To correct for this systematic observational artifact, a +1.8
km~s$^{-1}$ shift has been applied to all derived velocities presented
in this work (Table \ref{tbl-3}) and \citet{covey2005b}.

The width of the Gaussian fit in Fig. \ref{fig-radvel.offset} also
provides an indication of the level of accuracy with which we can
determine the velocities of targets ($\sigma \sim$ 2 km~s$^{-1}$).
The true distribution appears to have slightly broader wings than the
Gaussian fit.  We therefore consider the size of our true velocity
error as 1.5~$\sigma$, or $\sim$ 3 km~s$^{-1}$.

\subsection{Stellar Luminosities Derived from Photometry}\label{sec-photlums}

\begin{figure}
\figurenum{14}
\plotone{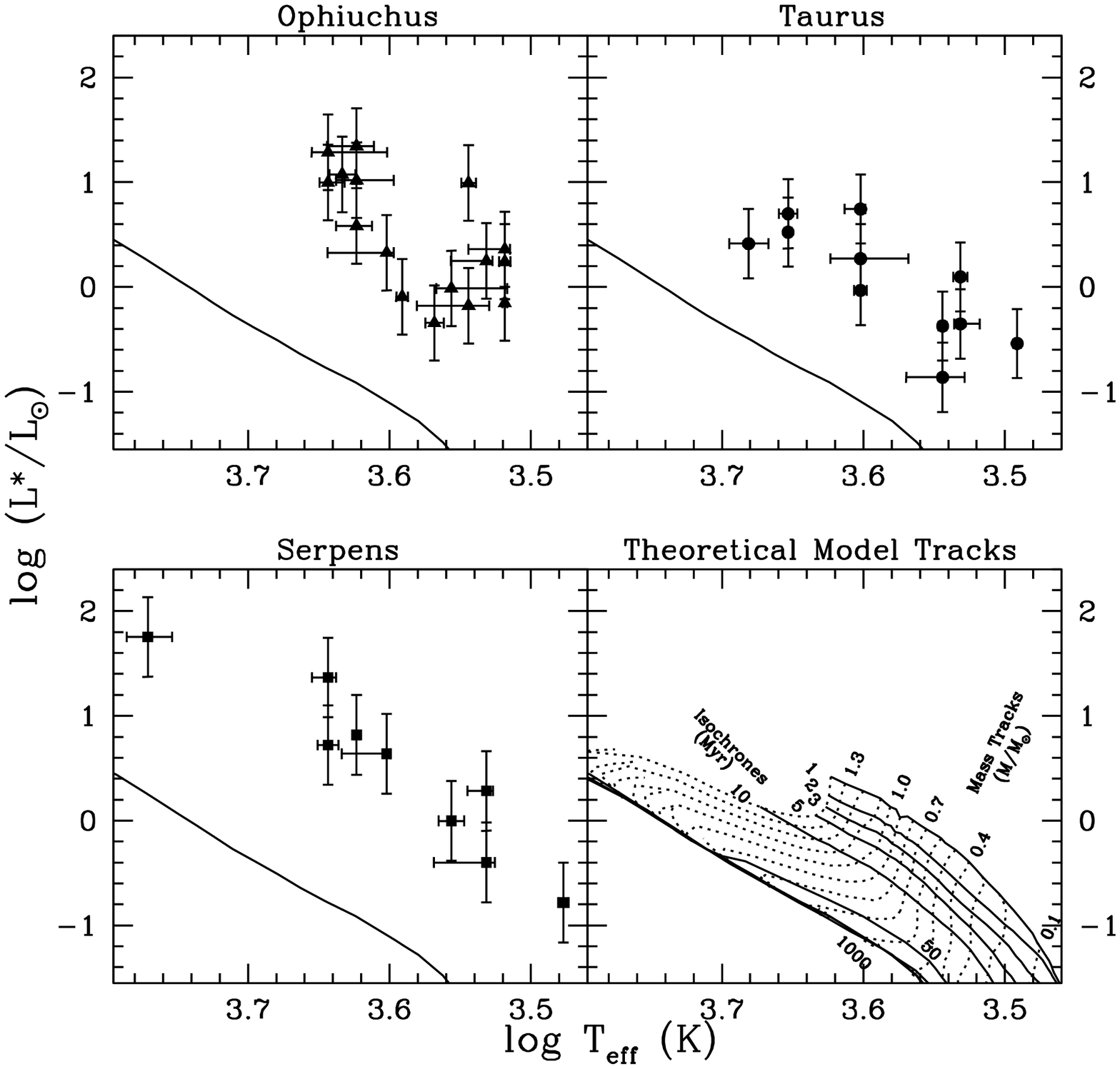}
\caption[HRD lstar]
{\label{fig-hrd.lstar}
Class I and flat-spectrum protostars are placed on the H-R diagram
using spectroscopically derived temperatures and photometrically
derived stellar luminosities without any accretion.  The Class I and
flat-spectrum YSOs in $\rho$ Oph (filled triangles, upper left panel),
Tau-Aur (filled circles, upper right panel), and Serpens (filled
squares, lower left panel) are broadly distributed (with 1~$\sigma$
error bars) above the theoretical main-sequence (solid line).
Theoretical mass tracks and isochrones are shown (lower right panel)
as dotted and solid lines, respectively \citep{baraffe1998}.}

\end{figure}

In addition to the surface gravities we measure directly from the
spectra, we derive stellar luminosities for our sources in Ophiuchus,
Taurus, and Serpens based on published H and K-band apparent
magnitudes \citep{eiroa1992,kenyon1995,barsony1997,kaas1999}.  Many of
our objects do not have J-band photometry, so we are unable to
de-redden these sources individually on a color-color diagram.
Instead, we estimate the extinction at K by de-reddening all our
objects to an intrinsic H--K color of 0.6, which is midway between the
end of the classical T-Tauri star locus \citep[H--K =
1.0,][]{meyer1997} and the average intrinsic color for late-type
dwarfs \citep[H--K $\approx$ 0.2, Table A5 of][]{kenyon1995}.  To
account for the presence of scattered light, we increase each of our
derived extinctions by 0.88 magnitudes (see paragraph below).
Absolute K magnitudes were computed from observed K magnitudes
corrected for extinction \citep[employing the extinction law
of][]{martin1990}, excess emission that is measured spectroscopically
though veiling in this work (see Table \ref{tbl-3}), and distance to
the star forming cloud \citep[Tau~=~140~pc, Oph~=~145~pc,
Ser~=~259~pc;][]{kenyon1995,dezeeuw1999,straizys1996}.  The stellar
luminosity is found by converting the absolute K magnitude to a V
magnitude adopting a V--K color and a bolometric correction
appropriate to a late-type dwarf having our derived effective
temperature \citep[Table A5,][]{kenyon1993}.  Using the photometric
luminosities and spectroscopic effective temperatures, these objects
are plotted on an observational H-R diagram, and show a spread in
T$_{\rm eff}$ and luminosity (Fig. \ref{fig-hrd.lstar}).  This stellar
luminosity should be representative of the PMS star and not include
any accretion luminosity.

The flux measured within a photometric aperture toward a Class I
object contains contributions from sources other than the direct light
from the extincted stellar photosphere.  These other sources of
emission include the surrounding accretion disk, hot dust in the inner
envelope and scattered light from holes or cavities in the envelope.
Indeed, observed high polarizations of many Class I sources in Taurus
suggest that the observed fluxes are dominated by scattered light
\citep{kenyon1993, whitney1997}.  In this case extinction estimates
based only on de-reddened colors are underestimates since they
represent the extinction to the scattering surface and not the
protostellar object itself.  In some cases this could lead to severe
underestimates in the source luminosity.  We compare our derived
extinction estimates to those derived by \citet{whitney1997} from a
more careful analysis using polarimetric observations and radiative
transfer models of Class I envelopes to derive more realistic
extinctions to individual protostellar objects in the Taurus cloud.
Seven of those objects are also in our sample, and we find that the
\citet{whitney1997} K-band extinctions are 0.88 magnitudes higher
(mean value) than the extinctions we compute from the H--K colors of
those same objects.  Therefore in an attempt to account for the
extinction produced by the presence of scattered light, we increase
the derived extinctions in all of our Class I and flat-spectrum
sources by 0.88 magnitudes.

The luminosity error bars depicted in Figure \ref{fig-hrd.lstar} are
the quadrature sum of uncertainties due to distance to the cloud,
observed photometry, slope of the extinction power law, the (assumed)
intrinsic (H--K) color, and the presence of scattered light.  The last
two quantities are highly uncertain due to the complexities of the
circumstellar environments typical of Class I objects.  The largest
uncertainty in our luminosity calculation arises from the intrinsic
colors of protostars (H--K = 0.6 $\pm 0.4$) which may not resemble
those of dwarfs.  To quantify the luminosity uncertainty due to
scattered light, we examine the uncertainty of the 0.88 magnitude
offset of the K band extinction values, the mean difference between
the extinctions derived by \citet{whitney1997} and us for the seven
objects in Taurus that we had in common.  We determine the uncertainty
of this offset from the standard deviation of the mean (0.39
magnitudes in A$_{\rm K}$).  Ascertaining appropriate extinctions and
disentangling the effects of scattering, source geometry and veiling
from disks and envelopes likely requires detailed modeling of the
envelopes of individual sources in order to derive meaningful
luminosities for the central embryonic stars.  In the absence of such
modeling one must regard our derived luminosities of Class I objects
as being highly uncertain.

\section{Discussion}

\subsection{H-R Diagrams}\label{sec-hrdiags}

\begin{figure}
\figurenum{15}
\plotone{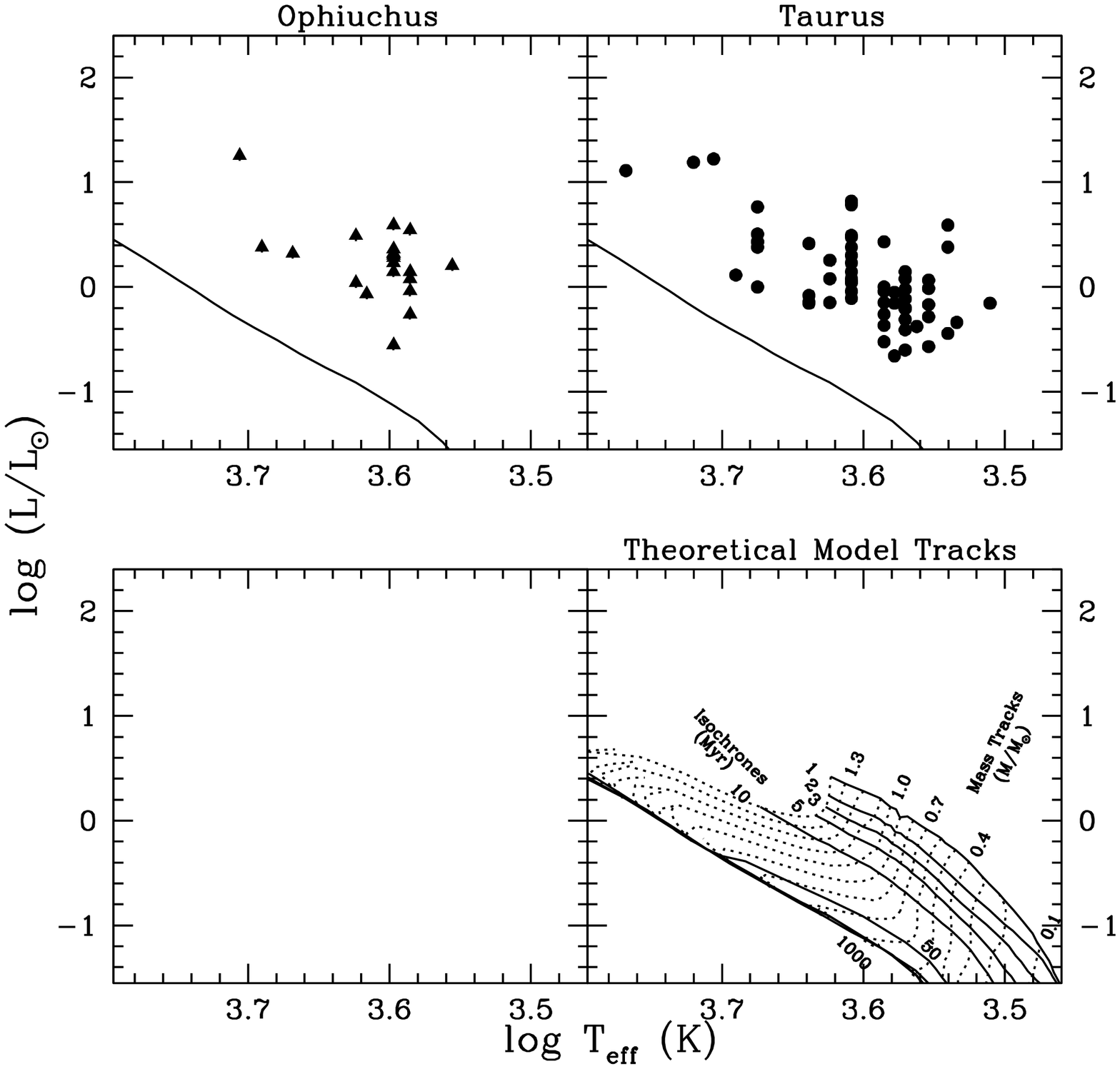}
\caption[HRDlstarclassII]
{\label{fig-hrd.lstar.classII}
Class II YSOs are placed on the H-R diagram based on literature
reported values of T$_{\rm eff}$ and L$_{\rm bol}$, as measured in a
spectroscopic survey of Ophiuchus \citep[filled triangles in upper
left panel,][]{luhman1999}, and a photometric survey of Taurus
\citep[filled circles in upper right panel,][]{kenyon1995}.  There is
a comparable scatter in luminosities of these sources as compared to
our Class I and flat-spectrum sources (fig. \ref{fig-hrd.lstar}),
corresponding to an equivalently large spread in ages based on
theoretical evolutionary model tracks \citep[lower right
panel,][]{baraffe1998}.}

\end{figure}

The placement of Class I and flat-spectrum sources in H-R diagrams is
one of the main goals of this study. Coupling our determinations of
T$_{\rm eff}$ and $\log~g$ with estimates of individual source
luminosities derived from existing photometric observations, we are
able to place our object sample on the H-R diagram.  We present the
individual H-R diagrams for Ophiuchus, Taurus and Serpens in Figure
\ref{fig-hrd.lstar}.  For comparison we also plot one set of
theoretical PMS evolutionary tracks \citet{baraffe1998} on an
accompanying H-R diagram.  We find the flat-spectrum / Class I stars
to lie in a relatively broad band, parallel to and well above the
main-sequence, but below the expected birth-line for YSOs
\citep{stahler1994}. In Figure \ref{fig-hrd.lstar.classII} we show the
locations on the H-R diagram of Class II sources from the same
clouds. Comparison of the two plots shows that Class I and
flat-spectrum central stars span a very similar range of luminosity
and effective temperature as Class II sources. For both types of
sources there is a relatively large scatter in their positions on the
H-R diagrams. These findings are similar to those derived from optical
spectroscopy of Class I sources in Taurus
\citep{kenyon1998,white2004}.

Classical PMS theory would suggest that Class I sources should appear
at or near the birth-line, however PMS models that include the effects
of accretion indicate that accreting stars evolve more quickly down
Hayashi tracks than non-accreting stars and should appear at lower
luminosities than non-accreting stars of the same age \citep{tout1999,
siess1999}. Thus the locations of Class I sources below the birthline
are consistent with expectations for accreting stars. However, because
Class I sources should be systematically younger than Class II objects
the Class I stars should on average be more luminous than the Class II
sources which does not appear to be the case for the objects we
observed. However as discussed below large uncertainties in both
derived source luminosities and the locations of theoretical tracks on
the H-R diagram may mask such an effect.

\begin{figure}
\figurenum{16}
\plotone{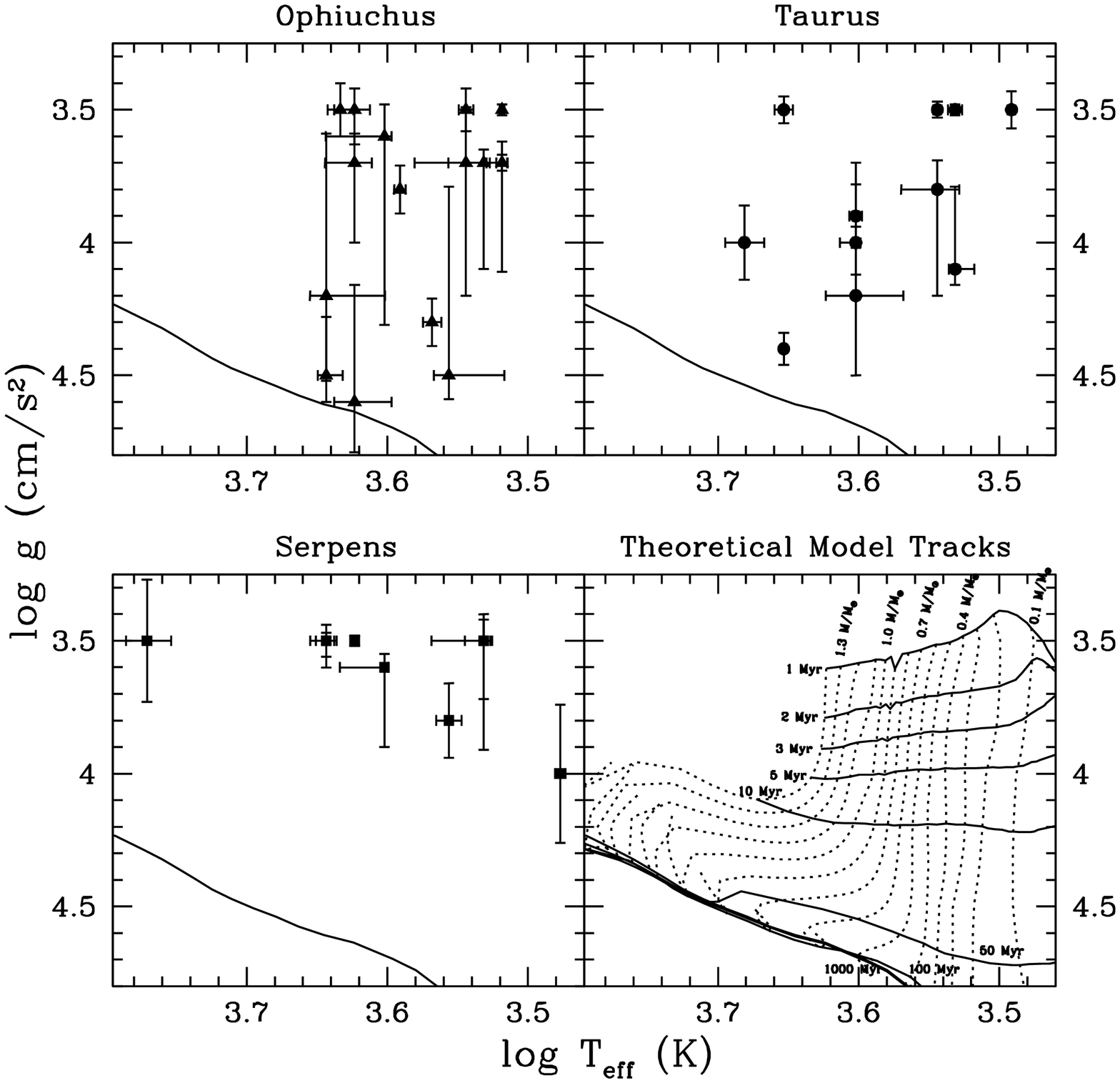}
\caption[HRD logg]
{\label{fig-hrd.logg} 
Class I and flat-spectrum protostars are plotted on the H-R diagram
based on effective temperatures and surface gravities derived from
spectroscopic fits (1~$\sigma$ error bars) to K-band photospheric
lines.  The Class I and flat-spectrum YSOs in $\rho$ Oph (filled
triangles, upper left panel), Tau-Aur (filled circles, upper right
panel), and Serpens (filled squares, lower left panel) are broadly
distributed within the search grid for temperature (T$_{\rm eff}$:
3000K -- 6000K) and gravity ($\log~g$: 3.5 -- 5.0), with many sources
residing at the low gravity edge, and two sources in $\rho$ Oph
(YLW~16A \& WL~19) with the highest gravities lying close to the
theoretical main-sequence (solid line).  Theoretical mass tracks and
isochrones are shown (lower right panel) as dotted and solid lines,
respectively \citep{baraffe1998}.}
\end{figure}

In Figure \ref{fig-hrd.logg}, we plot the surface gravities vs
effective temperatures for Class I and flat-spectrum sources in the
same clouds along with the predictions of theoretical models of PMS
evolution. The surface gravities we derive from spectroscopy are
correlated with the photometrically determined luminosities using
evolutionary model tracks to infer the stellar mass.  We find that our
protostellar sample are characterized by sub-giant surface gravities
that are significantly lower than those of main-sequence dwarfs. There
is a relatively large scatter in distribution of the Class I and
flat-spectrum surface gravities.

If we compare the positions of our YSOs on the H-R diagram and on the
surface gravity plots with the predictions of classical PMS theory we
find these YSOs appear to span an age range of $\sim$ 10$^7$ years,
similar to that of Class II objects. This is almost certainly an
unphysical result.  For example from statistical arguments the Class I
phase is thought to be only about 10$^5$ years in duration,
considerably shorter than the spread of sources on the H-R and surface
gravity diagrams would suggest.  Moreover, the observed
color-magnitude diagrams of slightly older ($>5$ Myr)
young clusters exhibit very well-defined pre-main sequences with
nearly coeval age spreads of typically less than 2--3 Myr
\citep{moitinho2001, preibisch2002}.  Together these facts suggest
that the observed spreads on the H-R diagrams for a very young stellar
population ($\sim$ 1--3 Myr) are a result of factors other than age,
factors that introduce systematic uncertainties in the locations of
stars on the H-R diagram.  These factors include intrinsic source
variability, uncertainty in extinction corrections, binarity
\citep[e.g.,][]{preibisch1999} as well as effects of accretion on the
positions of accreting stellar objects on the H-R diagram
\citep{tout1999, siess1999}.

Protostellar luminosities are dominated by scattering and disk
accretion and thus determination of the stellar component of the
luminosity from observed photometry is likely one of the largest
sources of uncertainty in the vertical positions of Class I and
flat-spectrum objects on the H-R diagram.  Errors in stellar
luminosities derived from near-IR photometry in our sources are
dominated by uncertainties in judging the intrinsic near-IR colors of
these embedded sources ($\S$ \ref{sec-photlums}).  Relying on the
assumption that these objects are radiating isotropically is
problematic since the stellar light can be partially blocked by disks
or scattered through cavities.  \citet{kenyon1990} find that the
stellar luminosity in T-Tauri stars can be underestimated by up to
$\sim$ 50\% (factor $\sim$2 in age) due to occultation by opaque disks
at various inclination angles.  Luminosities of Class I sources in
Taurus report uncertainties of $\sim$33\% - 50\% for objects whose
optical spectra are seen only in scattered light \citep{kenyon1998}.
The geometry of protostellar disks is clearly important to making
accurate measurements of the luminosity emanating from an embedded
source, and overestimates in the apparent age can result if this is
not accounted for.  This effect has been seen by others as well; the
H-R diagrams of \citet{white2004} show that Class I YSOs and YSOs
driving Herbig-Haro flows in Taurus-Auriga which have edge-on disks
have low luminosities making them to appear to have very high ages, on
the order of 10$^{8}$ yr.  Uncertainties in our spectroscopic
measurements of gravity stem from our fits to key photospheric lines
(see $\S$ \ref{sec-qualityfits}), whose shapes and depths are affected
by combinations of physical parameters (i.e. T$_{\rm eff}$, $\log~g$,
r$_{\rm K}$) that are difficult to completely separate in finite
signal-to-noise data.

A theoretical uncertainty in using H-R diagrams to measure masses and
ages lies in the reliability of the placement of evolutionary model
tracks which permit us to convert observables into fundamental
intrinsic properties.  The presence of accretion delays the formation
of a radiative core \citep{siess1997}, which offsets the isochrones
toward lower luminosities and higher gravities in cool (T$_{\rm eff}
<$ 5800K) YSOs \citep{tout1999}.  Depending on the accretion history,
and initial mass, overestimates in the ages vary between 30\% and a
factor of 5 when compared to models without accretion
\citep[i.e.][]{baraffe1998}.

It is quite likely therefore that the large spreads in the stellar
luminosities and gravities that we find for our Class I and
flat-spectrum sources are not due to real spreads in age, but
represent uncertainties in the intrinsic colors, extinctions, and
emitting geometry of each star, along with observational uncertainties
in their spectra.  Therefore differences in the stellar luminosities
of Class I / flat-spectrum and Class II sources are extremely
difficult to detect in observational H-R diagrams and comparisons of
derived luminosities and ages between these objects are not likely
meaningful.  Moreover, when the observational uncertainties are
coupled with uncertainties in the PMS evolutionary tracks it is
evident that Class I and flat-spectrum ages can not be confidently
extracted from H-R diagrams.

\subsection{CO Emission}\label{sec-coemis}

Near-IR CO emission has been observed in a number of YSOs
\citep{scoville1983,geballe1987,carr1989}, although it is less common
in low luminosity objects \citep[e.g.][]{greene1996,luhman1998}.  CO
emission has been used as a probe to study the kinematics of the inner
disk region around YSOs \citep{najita1996}, since relatively warm
(T$_{\rm ex}$ = 2500K -- 4500K) and dense (n $> 10^{9}$ cm$^{-3}$)
neutral gas is needed to collisionally excite CO, where the origin of
the emission may likely be from a temperature inversion in the inner
disk atmosphere.  While another source of this emission could be
outflows of neutral gas in winds \citep[e.g. SVS~13,][]{carr1992},
most CO overtone emission fits profiles of rapid Keplerian rotation
(up to 400 km~s$^{-1}$) originating from an inner disk
\citep{dent1991,carr1993,chandler1995}.

The rotation profile of a disk is distinctly double peaked in velocity
space \citep{carr1993} because different disk regions have different
Keplerian velocities and different CO emission strengths.  When this
profile is convolved with the first overtone CO bandhead and R branch
lines, the result is an asymmetric profile having a blue wing, a
shoulder, and a peak redward of the bandhead rest velocity
\citep{najita1996}.  In contrast, a stellar rotation profile
\citep{gray1992} has a single symmetric peak centered at the rest
velocity, with the higher velocity components contributing less as the
projected emitting area decreases outward toward the limb.

We detect CO emission in 8 of our sources (15\% of our sample), which
is accompanied by Br~$\gamma$ emission in all cases, and Na emission
at 2.2062 $\mu$m \& 2.2090 $\mu$m in 5 cases.  Except for 2 sources
(EC~129 \& EC~38, see Appendix \ref{sec-indivsources}), the broadened CO
emission ($\le$~100 km~s$^{-1}$) shows a velocity structure that is
consistent with spectral models of Keplerian rotation \citep{carr1993,
najita1996} from an optically thin disk.  Two sources have both CO and
Br~$\gamma$ absorption superposed on corresponding emission features
in these two lines.

\subsection{Brackett $\gamma$}\label{sec-brgemis}

\begin{figure}
\figurenum{17}
\plotone{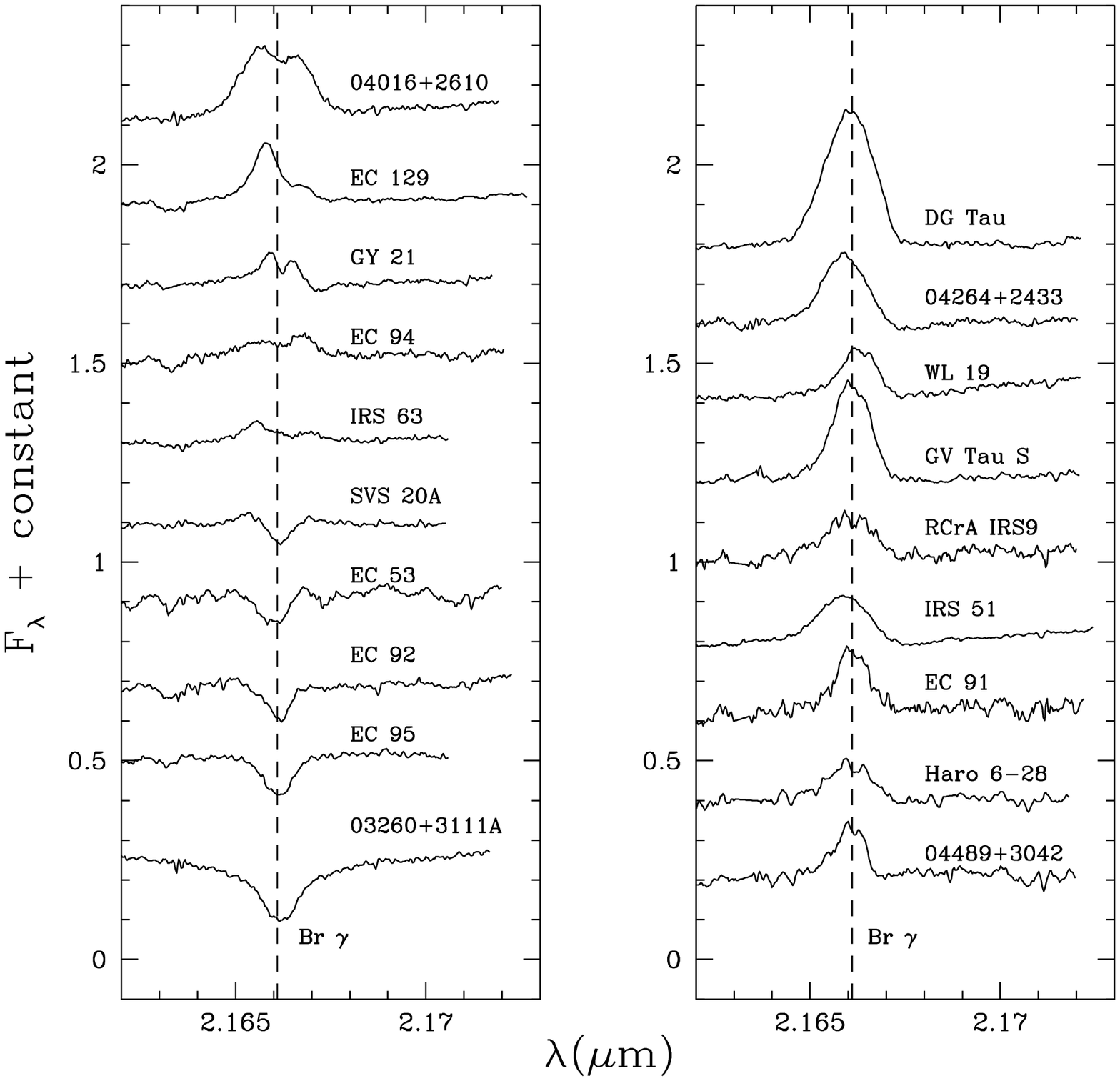}
\caption[brg.spec1] 
{\label{fig-brg.spec1} Br~$\gamma$ emission and/or absorption is
evident in the majority of the protostars in our sample.  The spectra
in both panels have each been individually shifted to remove the
effects of all radial velocity motions measured in their photospheric
lines in other orders (Table \ref{tbl-3}).  The vertical dashed line
shows the rest velocity of the Br~$\gamma$ line.  In most cases, the
absorption peak is coincident with the rest velocity, indicating a
photospheric origin.  The right panel shows a hint of Br~$\gamma$
absorption in several sources that all have strong and broad emission
profiles, with some absorption coincident with the rest (photospheric)
velocity.}

\end{figure}

\begin{figure}
\figurenum{18}
\plotone{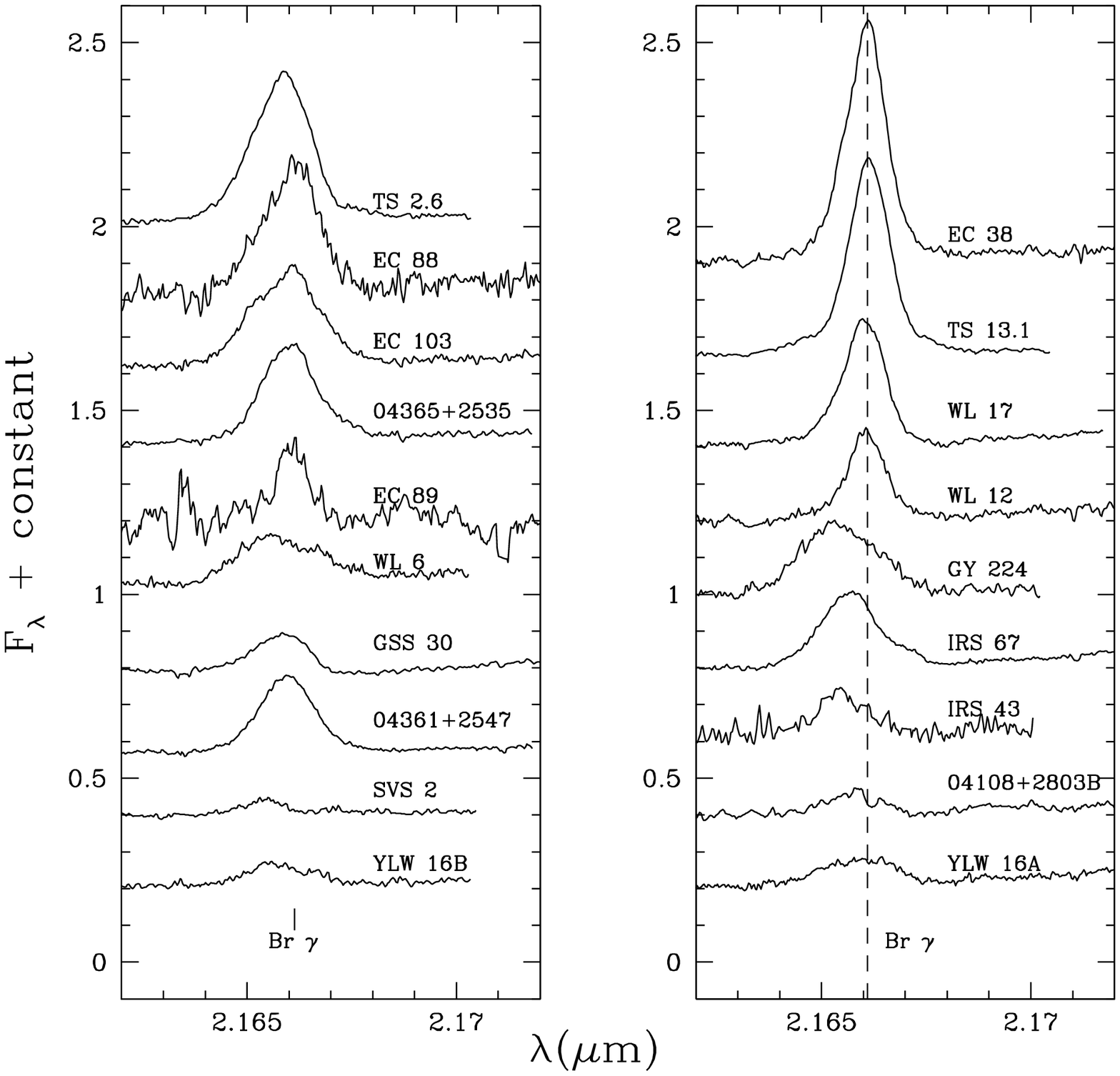}
\caption[brg.spec2] 
{\label{fig-brg.spec2}
Ten of 11 protostars in our sample which did not not show photospheric
lines needed for our spectral analysis have Br~$\gamma$ emission (left
panel) over a range in line widths (FWHM 95 -- 300 km~s$^{-1}$).  The
right panel shows Br~$\gamma$ in sources where we have measured the
stellar radial velocity from photospheric lines present in other
orders.  These spectra in the right panel have each been individually
shifted to remove the effects of all radial velocity motions seen in
their photospheric lines.  The vertical dashed line shows the rest
velocity of the Br~$\gamma$ line.  GY~224, IRS~67, and IRS~43 stand out
as having distinctly blue-shifted Br~$\gamma$ emission.}
\end{figure}

We detect HI Br~$\gamma$ emission (2.1661 $\mu$m) in 34 of 52 YSOs
observed in our sample (65\%, shown in Figs. \ref{fig-brg.spec1} \&
\ref{fig-brg.spec2}).  In general, the emission profiles are broad
($\sim$95 -- 320 km~s$^{-1}$, FWHM), consistent with what is seen
in low luminosity embedded YSOs and active T-Tauri stars
\citep{najita1996}.  In 4 cases (i.e. EC~129, GY~224, IRS~67, IRS~43),
the emission is significantly blue-shifted relative to the stellar
systemic velocity, indicative of an outflow.  These sources are among
those that are marked with dashed vertical line indicating the objects
where we measure photospheric radial velocities
(Figs. \ref{fig-brg.spec1} \& \ref{fig-brg.spec2}).

In 6 cases, there is accompanying Br~$\gamma$ absorption that is
superposed near the peak of the emission profile (left panel of
Fig. \ref{fig-brg.spec1}), but red-shifted by $\sim$15 -- 65
km~s$^{-1}$(i.e. 04016+2610, EC~129, GY~21, EC~94, IRS~63, SVS~20A).
Emission with blue-shifted absorption, which would be explained by an
expanding wind \citep{najita1996}, is absent in our sample.  Instead,
the red-shifted absorption minimum is characteristic of material
in-falling from the disk onto the photosphere \citep{muzerolle1998},
perhaps along magnetic accretion columns in cases of greater velocity
red-shifts where in-falling material can reach free fall velocities.
Inverse P-Cygni profiles in classical T-Tauri stars indicative of
in-fall are ubiquitous \citep{edwards1994}, and have been explained by
free falling gas along the magnetosphere
\citep{hartmann1994a,hartmann1994b}.  The modest red-shifts (v$_{\rm
r} <$ 65 km~s$^{-1}$) we observe in these Br~$\gamma$ absorption
components (relative to the emission peak) suggests that accretion may
be occurring through the inner disk in this stage before it could
become disrupted and truncated in a later evolutionary stage
\citep{bertout1988}.

Four YSOs have Br~$\gamma$ absorption without the presence of any
emission.  The strongest and broadest of these is 03260+3111A, whose
equivalent width (3.4~${\rm \AA}$) and broadened line wings are
consistent with Br~$\gamma$ absorption we see in late G-type dwarf
standards, confirming its relatively warm T$_{\rm eff}$ (5600K)
derived from spectral fits to photospheric lines in the Na and Mg/Al
intervals.  Only one other source is warm enough (SVS~20A, see Appendix
\ref{sec-indivsources}), to allow a photospheric origin for the
absorption.  For the other 3 sources that show only Br~$\gamma$ in
absorption (EC~53, EC~92, EC~95), we see equivalent widths that are
significantly greater than what is observed in the MK standards of
comparable spectral-type.  The likely explanation for this is that
either the Br~$\gamma$ absorption arises from a non-photospheric
mechanism \citep{muzerolle1998}, or else these are binary systems with
both early and late-type components. If the latter is true, then the
early-type star must cause significant veiling of the K-band
photospheric features of the late-type star in each system, by nature
of their relative temperatures and the larger radii of the early-type
stars.  This does not apply for EC~92 and EC~95 which have veiling
r$_{\rm K}$ = 0.0 and 0.1, respectively.  In the case of EC~53 which
has veiling r$_{\rm K}$ = 1.8, an earlier spectral-type binary
component that is warm enough to have photospheric Br~$\gamma$
absorption lines (i.e. T$_{\rm eff}$ $>$ 4200K) would mask the lines
(r$_{\rm K} > 5$) of the cooler star (T$_{\rm eff}$ = 3400K).
Therefore we conclude that the Br $\gamma$ absorptions of EC~53, EC~92
and EC~95 most likely arise from another mechanism.  The radial
velocities of all the Br~$\gamma$ absorption lines lie at the stellar
systemic velocity to within one resolution element of our observations
(17 km~s$^{-1}$).

\subsection{Excess Continuum Emission}\label{sec-cont.emis}

K-band veiling traces thermal emission from circumstellar material.
Moderate amounts of veiling ($\rm r_{\rm K} \lesssim 2$) can be
produced by circumstellar disks, depending on their structures,
activities, and geometries. Circumstellar envelopes are usually
responsible for larger amounts of veiling
\citep[see][]{greene1996}. Young stars with large r$_{\rm K}$ values
have more warm circumstellar material than less veiled ones. K-band
veiling may also be a good indicator of mass accretion because matter
close to a young star is attracted to it by gravity.

We now use r$_{\rm K}$ measurements to investigate whether the Class I
and flat-spectrum objects in our sample have more warm circumstellar
material than Class II and T-Tauri stars. We start by compiling mean
veiling values for each of the 3 major star forming regions which we
have studied. We compute these values from the r$_{\rm K}$ results
presented in Table 3, but we also correct for the objects for which we
were not able to determine physical parameters. Eight of these objects
(2 in Tau-Aur, 3 in $\rho$ Oph, and 3 in Serpens) were observed with
high signal-to-noise, so we assign them the value r$_{\rm K}$ = 4.0.
We determined this value to be their likely lower limits using our
observed veiling / rotation completeness limit assuming a typical
rotation velocity $v\sin~i$ = 35 km~s$^{-1}$ (see Fig.
\ref{fig-vsini.veil}). The other 2 objects (CRBR~85 in Oph and EC~89
in Ser) show weak Na and CO absorption features (too weak for
analysis) in moderate signal-to-noise spectra, so we assign them both
r$_{\rm K}$ = 1.5, typical for the object sample.

With these corrections, our entire Class I and flat-spectrum object
sample has a mean veiling $<\rm r_{\rm K}>$ = 1.83. The objects in
Tau-Aur and Ser both have $<\rm r_{\rm K}>$ = 1.54 ($\pm$1.21 for
Tau-Aur and $\pm$1.49 for Ser), while the $\rho$ Oph objects are
characterized by $<\rm r_{\rm K}>$ = 2.21 $\pm$1.35. We evaluated the
likelihood that objects in these different regions have statistically
significant veiling differences by performing KS tests on their r$_{\rm K}$
values. We found that the Ser and Tau-Aur objects are likely to be
drawn from the same veiling distribution (88\% probability), while the
Tau-Aur and $\rho$ Oph probability was 18\% and the Ser and Oph
probability was 24\%. Thus all the objects in the 3 major star forming
regions have similar veiling values, with $\rho$ Oph and Tau-Aur
showing some evidence for being different (at the 1.3~$\sigma$
level). We infer similar conclusions about the similarity of
circumstellar material around these objects.

Next we compare these results to the veilings of Class II YSOs in
order to evaluate whether objects in our study have significantly more
circumstellar material. Of the 3 major star formation regions in our
study, only the $\rho$ Oph region has a significant number of Class II
YSOs for which r$_{\rm K}$ values have been determined. We compute
that the mean veiling of Class II objects in $\rho$ Oph is $<\rm
r_{\rm K}>$ = 0.94 $\pm$1.03 from data presented by
\citet{doppmann2003b} (10 objects) and \citet{luhman1999} (15
objects). This is about half that of the $\rho$ Oph objects in our
Class I and flat-spectrum sample.  Furthermore, the KS test reveals
that the r$_{\rm K}$ values of the Class I / flat-spectrum and the
Class II samples in $\rho$ Oph have less than a 1\% chance of being
drawn from the same parent population.  Therefore we conclude that
Class I and flat-spectrum objects have significantly more veiling than
Class II YSOs. The Class I and flat-spectrum objects must have more
warm circumstellar material, making it likely that they are in a more
embedded and younger evolutionary state.

\begin{figure}
\figurenum{19}
\plotone{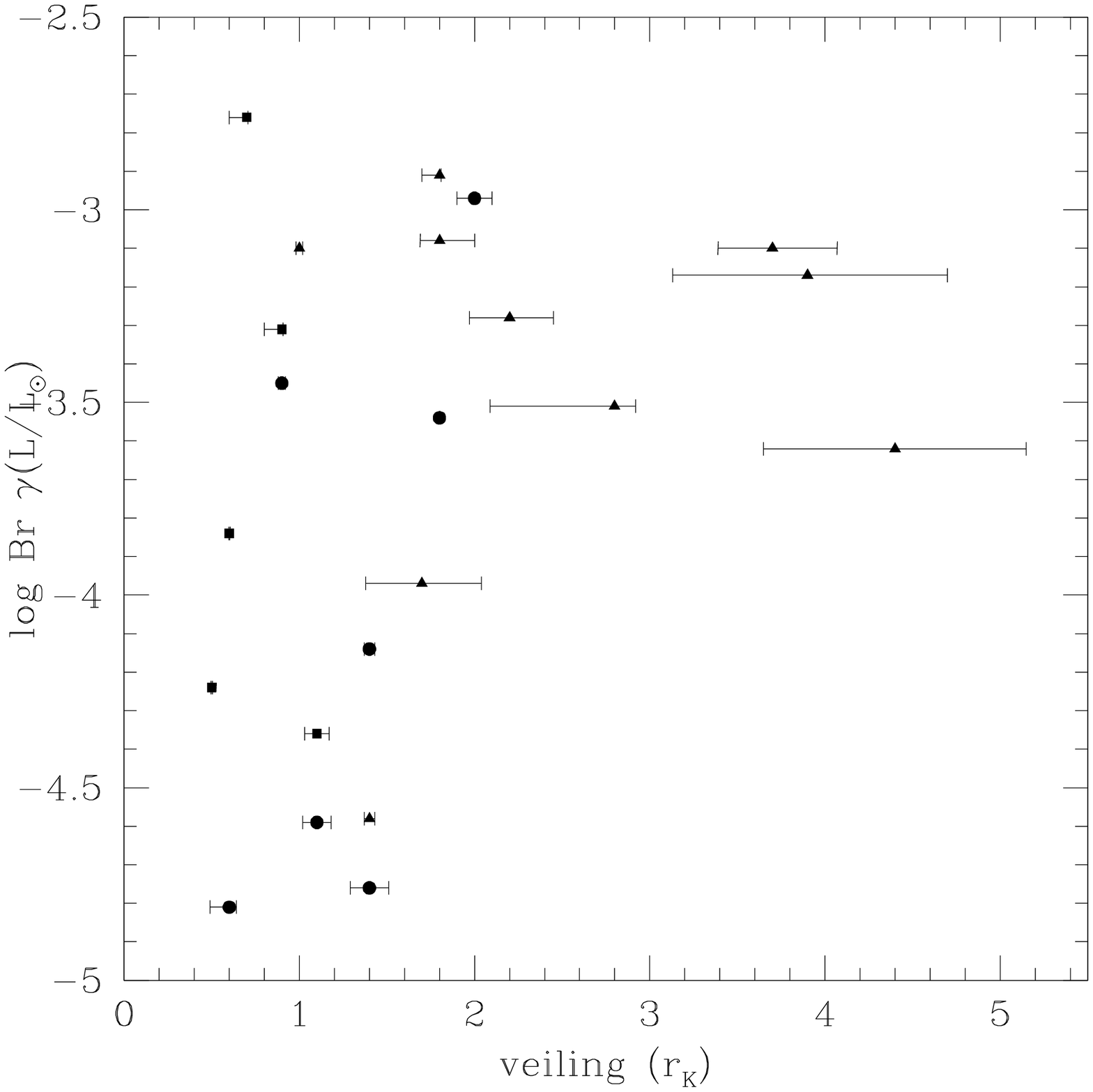}
\caption[brg-veil]
{\label{fig-brg.veil}
Brackett$~\gamma$ luminosity is computed from the equivalent width of
the emission feature at 2.1661 $\mu$m in our sample for the 22 sources
where we also have measured veiling by spectroscopic analysis of
photospheric absorption lines from other orders.  These protostars
showing Br~$\gamma$ emission, 10 in Ophiuchus (triangles), 7 in Taurus
(circles), and 5 in Serpens (squares) have had their Br~$\gamma$
equivalent widths converted to luminosities using H and K photometry
in the literature, and correcting for extinction in the presence of
scattered light (see $\S$ \ref{sec-photlums}).  The general trend of
increased veiling with stronger Br~$\gamma$ luminosity, evident here,
is expected as both these quantities are believed to trace accretion
activity, with the sources in Ophiuchus appearing to be most active of
the three regions.}

\end{figure}

\citet{muzerolle1998b} have shown that HI Br~$\gamma$ emission is
well-correlated with mass accretion in T-Tauri stars, and we now
evaluate whether r$_{\rm K}$ and Br~$\gamma$ emission are correlated
in our Class I and flat-spectrum sample.  In Figure \ref{fig-brg.veil}
we plot Br~$\gamma$ line luminosity (after correcting for extinction
and distance) for sources where we also were able to measure r$_{\rm
K}$.  These two parameters are positively correlated, but Fig.
\ref{fig-brg.veil} shows too much scatter (perhaps due to uncertain
extinction corrections or asymmetrically distributed circumstellar
material) to confirm that both r$_{\rm K}$ and Br~$\gamma$ emission
are reliable mass accretion diagnostics for Class I and flat-spectrum
YSOs.  Using our measured Br~$\gamma$ luminosities, the
\citet{muzerolle1998b} relation suggests that our $\rho$ Oph sources
have accretion luminosities 10$^{-1} < L_{\rm acc}/L_{\odot} <
10^{0}$, while Taurus and Serpens have about an order of magnitude
less.  However, we note that many of the objects in our sample have
been observed to have significantly higher bolometric (accretion)
luminosities \citep{wilking1989}, and the \citet{muzerolle1998b}
relation was derived for Class II objects with low accretion
luminosities.  Therefore the absolute accretion luminosities derived
from Br~$\gamma$ fluxes are likely to be underestimates, although the
relative difference between $\rho$ Oph and the other regions is
interesting and may be real.

\subsection{The Natures of Class I and Flat-Spectrum Sources}\label{sec-class1.natures}

The above discussion has demonstrated that analysis of spectroscopic
features provides much more detailed information on the natures of
YSOs than can be ascertained from SED shapes alone.  We now examine
the extent to which the aggregate of information derived from our
spectra provides insight into the true physical natures and
evolutionary status of Class I and flat-spectrum YSOs. As part of this
effort, we also assess whether this new information supports the
established paradigm that Class I objects are protostars near the ends
of their bulk mass accretion phases that then evolve to Class II or
Class III YSOs.

Our analysis has shown that Class I and flat-spectrum YSOs have some
astrophysical properties which are similar and some which are
considerably different from Class II and III YSOs. As noted in $\S$
\ref{sec-hrdiags}, our Class I and flat-spectrum sample have effective
temperatures, surface gravities, and stellar luminosities which are
similar to those of Class II PMS stars. This is not completely
unexpected given the relatively large uncertainties in stellar
luminosities determined for Class I and flat-spectrum objects ($\S$
\ref{sec-photlums}; see also Figure \ref{fig-hrd.lstar}), and the
stellar cores of steadily accreting protostars are expected to be in
near hydrostatic equilibrium with radii and central temperatures
similar to T-Tauri stars \citep{stahler1980}.

However, other intrinsic astrophysical properties of Class I and
flat-spectrum objects do differ significantly from Class II or Class
III PMS stars.  In $\S$ \ref{sec-cont.emis}, we determined that the
K-band veilings of $\rho$ Oph Class I and flat-spectrum YSOs are
significantly higher than those of $\rho$ Oph Class II objects,
indicating more circumstellar material and higher mass accretion in
Class I and flat-spectrum objects.  Also, our Class I and
flat-spectrum sample has significantly higher $v$ sin $i$ rotation
velocities and angular momenta than Class II and T-Tauri PMS stars the
Tau-Aur and $\rho$ Oph regions.  If a young star is magnetically
locked to its disk via its magnetic field, popular magnetospheric
accretion models predict that the disk coupling radius is proportional
to $\dot{M_*}^{-2/7}$, where $\dot{M_*}$ is the mass accretion rate
onto the star \citep{koenigl1991,shu1994}.  This means that YSOs with
higher accretion rates should be rotating faster than low accretion
rate objects, with angular velocities $\omega \propto
\dot{M_*}^{0.43}$ \citep[for a full discussion see][]{covey2005a}.

In summary, all fundamental astrophysical properties derived in this
study appear to support the paradigm that Class I and flat-spectrum
YSOs are accreting protostars. In this respect our results appear to
differ from those of \citet{white2004} who found that most
optically-selected Class I YSOs in Tau-Aur have mass accretion rates,
accretion luminosities, and rotation velocities that are not
significantly different from Class II and T-Tauri stars in the same
region.  Furthermore, \citet{white2004} conclude that the similarity
of Class I and Class II / T-Tauri spectral-types and stellar
luminosities imply that these objects have similar ages and
masses. Therefore \citet{white2004} argue that Class I and II YSOs are
not in different evolutionary states and that Class I objects are no
longer in their main mass accretion phases of evolution.  We note that
we argued in $\S$ \ref{sec-hrdiags} that conventional PMS models
cannot yield ages of Class I and flat-spectrum YSOs because they lack
accretion physics.

In an attempt to reconcile these results, we first investigate whether
the optical spectroscopy analysis technique of \citet{white2004}
yields the same parameters as our analysis for objects observed in
both studies.  Eight objects were observed in both studies, and we
derive astrophysical parameters for all of them.  The six
sources with astrophysical properties derived in both studies are
04016+2610, 04158+2805, DG~Tau, GV~Tau, Haro~6-28, and
04489+3042. Both studies derive similar effective temperatures
\citep[within 350K; converted from the spectral-type designations
of][]{white2004} for all 6 objects except DG~Tau, which
\citet{white2004} find to be significantly earlier (about 650K warmer)
than we do (Table \ref{tbl-3}). The high optical and IR veilings of
this object may contribute to uncertainties in the spectral-type and
effective temperature derived for this object. The $v\sin~i$ rotation
velocities of 5 of these 6 objects also agree well (to several
km~s$^{-1}$), with the exception of 04016+2610. \citet{white2004}
derive $v\sin~i \leq 15$ km~s$^{-1}$ for this object, while we report
46 km~s$^{-1}$. We suggest that the very low signal-to-noise in the
optical spectrum (S/N $\sim$ 2) may be responsible for this
discrepancy, and it is likely that our higher signal-to-noise spectrum
(S/N $\simeq$ 240) yields the less uncertain value. Overall it appears
that both studies yield similar astrophysical parameters for common
sources.

The differences in the conclusions between the two studies may be due
to differences in the underlying physical properties of the IR and
optically-selected samples that were analyzed.  \citet{white2004} find
that Class I YSOs do not have higher rotation velocities or higher
mass accretion rates (as inferred from veiling and luminosity
measurements) than Class II objects.  They do not derive astrophysical
parameters for two of the eight objects common in both studies,
L1551~IRS5 and 04295+2251.  Interestingly, we find that the $v\sin~i$
rotation velocities of these objects are 31 and 51 km~s$^{-1}$,
considerably higher than the median rotation velocity of 19.8
km~s$^{-1}$ of the \citet{white2004} Class I sample (see their Table
3).  The five Class I or flat-spectrum Tau-Aur YSOs in our sample
which were not observed by \citet{white2004} are not optically
visible, and we find 4 of the 5 to be highly veiled r$_{\rm K} > 1$,
with 2 of those veiled to the point where they have no discernible
absorption features (r$_{\rm K} \sim 4$; see Figure
\ref{fig-vsini.veil}).  This suggests that optically-selected Class I
YSOs may have less veiling and less accretion luminosity than
IR-selected ones.  Also, Tau-Aur YSOs have been known to have lower
bolometric luminosities \citep{kenyon1990b}, and presumably lower mass
accretion luminosities and rates than those in the $\rho$ Oph star
forming region \citep{wilking1989}.  Finally, we note that
\citet{white2004} did not account for the fact that several of their
Class I objects (and a few of their Class II ones) were too veiled to
determine their astrophysical parameters, so they did not compute mass
accretion rates or rotation velocities for these objects.  Therefore
the median mass accretion rates and rotation velocities computed in
that study are likely biased by these omissions.

The above discussion suggests that Class I sources span a wide range
of stellar and circumstellar properties, from actively accreting to
relatively quiescent objects.  We do believe that our large sample of
IR-selected Class I and flat-spectrum YSOs from several star
forming regions does give the most accurate picture to date of their
true natures as accreting protostars.

\section{Summary}

We have acquired high resolution near-IR spectra of a sample of 52
Class I and flat-spectrum YSOs located in the Tau-Aur, $\rho$ Oph,
Ser, and other nearby dark clouds. We have derived their fundamental
astrophysical properties via fitting their spectra to theoretical
spectra synthesized from the NEXTGEN \citep{hauschildt1999} PMS models
and make the following conclusions:

\begin{enumerate}

\item We detect absorption lines in 41 of the 52 objects.  This is a
much higher number and fraction than earlier studies because of the
high sensitivity and spectral resolution of the Keck II telescope and
its NIRSPEC spectrograph.  Twenty-three of the objects show 2.2235
$\mu$m H$_2$ emission lines.

\item Thirty-four objects show HI Br~$\gamma$ emission lines, and 6 of
these show Br~$\gamma$ absorption as well. The absorptions are
modestly red-shifted in all cases, indicating accretion in those
objects.

\item $\Delta v = 2$ CO emission is seen in 15\% of the objects,
indicating that CO emission and active circumstellar disks are more
common than expected from earlier studies.

\item The absorption spectra are fit well by the synthetic spectra,
indicating that the objects have effective temperatures and surface 
gravities which are consistent with PMS stars. The spectral fitting
procedure also yields rotation velocities, radial velocities, and
K-band veilings.

\item The objects show a wide range of effective temperatures and
stellar luminosities, and these values overlap with those of Class II
and T-Tauri stars when placed in H-R diagrams. The Class I and
flat-spectrum objects have low masses which span the same range as
Class II and T-Tauri stars, but significant observational and theoretical
uncertainties preclude deriving meaningful and accurate luminosities
and ages for protostellar sources from placement on the H-R diagram
and comparison with conventional PMS models.

\item We find that the mean K-band veiling of Class I and
flat-spectrum YSOs in $\rho$ Oph is significantly higher than for
Class II objects in this region. This indicates that the Class I and
flat-spectrum objects are accreting at higher rates, which is
bolstered by an analysis of their angular momentum properties in a
companion paper \citep{covey2005a}. We also find that K-band veilings
vary somewhat between Class I and flat-spectrum YSOs in different
regions, with the $\rho$ Oph objects having the highest veiling
values, $<\rm r_{\rm K}> = 2.21$.

\item We conclude that Class I and flat-spectrum YSOs are indeed
actively accreting protostars, although they span a wide range of
accretion activity. High resolution near-IR spectra provide
sufficiently detailed information to diagnose the stellar and
circumstellar properties of nearly any Class I or flat-spectrum YSO.

\end{enumerate}

\acknowledgments

The authors wish to recognize and acknowledge the very significant
cultural role and reverence that the summit of Mauna Kea has always
had within the indigenous Hawaiian community.  We are most fortunate
to have the opportunity to conduct observations from this mountain.
We also thank the staff of the Keck Observatory for excellent support
in the operation of the telescope and NIRSPEC.  We thank Russel White
for his careful reading of this manuscript and valuable comments.  We
thank Dan Jaffe and Amanda Kaas for helpful discussions.  We are
grateful to the referee, Barbara Whitney, whose insightful comments
helped improve the manuscript.  We acknowledge support from NASA's
Origins of Solar Systems program via UPN 344-39-00-09.
G.W.D. gratefully acknowledges support from the National Research
Council.  K.R.C gratefully acknowledges the support of NASA grant
80-0273.  This research utilizes NASA's Astrophysics Data System
Bibliographic Services, the SIMBAD database, operated at CDS,
Strasbourg, France, and the VizieR database of astronomical catalogs
\citep{ochsenbein2000}.






\appendix

\section{Notes on Individual Sources}\label{sec-indivsources}

EC~38:

Of all our sources which show CO overtone emission, EC~38 is the only
one with an extremely narrow CO emission profile ($\sim$30
km~s$^{-1}$), perhaps the most narrow observed CO overtone emission
profile seen in any embedded source.  Furthermore, there are two
resolved bandhead peaks (separated by 64 km~s$^{-1}$), with evidence
of beating in the R branch lines.  A conical high velocity neutral
wind in the plane of the sky viewed at low inclination angles
\citep{chandler1995} might explain the well-separated blue and
red-shifted bandhead components.  However, narrow CO disk emission
lines (40--50 km~s$^{-1}$ wide) have only been observed in one other
YSO (SVS~13), but are too narrow to fit emission models for either a
circumstellar disk or outflow \citep{carr1992}, unless there is
rotational coupling between the star and inner disk or if CO
originates from a lower velocity flow.  Strong Br~$\gamma$ emission is
also observed in this source. A model that includes in-falling
material from a truncated disk accompanied by a outflowing wind might
best describe these observations.

SVS~20A:

We detect broad Br~$\gamma$ and CO emission, along with Br~$\gamma$
absorption.  A marginal detection of weak CO absorption may also be
present with the CO emission.  The relatively warm T$_{\rm eff}$
(5900K) we derive from photospheric lines in the Na and Mg/Al
intervals and the presence of Br~$\gamma$ absorption at the same
radial velocity, argues that the origin of the Br~$\gamma$ absorption
is photospheric.  If CO absorption is also present, it must originate
in a disk or a wind, since photospheric CO is dissociated in warmer
stars.  If the CO feature is due to emission only, then its broadened
double peaked profile suggests that it may originate in a wind at a
high inclination angle \citep{chandler1995}. The radial velocity of CO
is blue-shifted ($\ge$70 km~s$^{-1}$) relative to the systemic stellar
velocity, in support of an outflow origin for the CO emission.  This
object is the warmest and most luminous protostar that we measure
stellar properties for.  Without making a correction for scattered
light (see $\S$ \ref{sec-photlums}), we derive a stellar luminosity of 24.7
L$_{\odot}$, based on the apparent flux at K from the North component
\citep[m$_k$=8.7,][]{eiroa1987}, adopting an H--K color of 2.16
observed in the total SVS~20 system \citep{kaas1999}.

EC~129:

This source displays emission and absorption components in both CO and
Br~$\gamma$ lines. EC~129 is cool enough (T$_{\rm eff}$= 4400K) for CO
to exist in the photosphere, though the blue-shifted radial velocity
($\sim$55 km~s$^{-1}$) relative to the systemic velocity of the star,
argues against the presence of photospheric CO.  Alternatively, CO can
exist in absorption in the disk if the effective temperature of the
disk increases faster than the surface temperature, favored in cases
of high mass accretion \citep{calvet1991}.  The CO emission
(blue-shifted even more than the absorption) could then be formed in
an outflow wind, while the red-shifted ($\sim$50 km~s$^{-1}$)
Br~$\gamma$ absorption line indicates in-falling material and
accretion.

DG~Tau:  

We do not detect any sign of CO overtone emission in our Nov. 2001
spectrum, as has been reported in the past
\citep[e.g.][]{chandler1995}, though this source is known to have
variable emission on week long time scales (reported in private
communication).  We find a good fit to the absorption lines, deriving
an effective temperature of 4000K and $\log~g$ = 4.0, with little
$v\sin~i$ rotation (24 km~s$^{-1}$), but significant veiling (r$_{\rm
K}$ = 2.0).

L1551~IRS5:

This prototypical bipolar molecular outflow has been well-studied for
over 20 years \citep[see review by][]{fridlund1997}.  Though we did
not detect HI Br~$\gamma$ emission or absorption in November 2001,
this source has displayed a P-Cygni H$\alpha$ profile early-on,
suggesting that it is an FU Orionis class YSO \citep{mundt1985}.  We
detect strong H$_2$ emission with a complex velocity structure at
2.2235 $\mu$m, indicating kinematically active shocked molecular gas.
Based on K-band absorption lines of Na, Mg, and Al, we derive an
effective temperature of 4800K, consistent with earlier estimates of
its spectral type from spectroscopy and photometry
\citep{carr1987,kenyon1995}.  We also derive a slow rotation rate
($v\sin~i$ = 31 km s$^{-1}$) and moderate veiling (r$_{\rm K}$ = 1.0)
in this embedded binary source.


\section{Atlas of Spectra}\label{sec-spectra}


\begin{figure}
\figurenum{1}
\epsscale{0.85}
\plotone{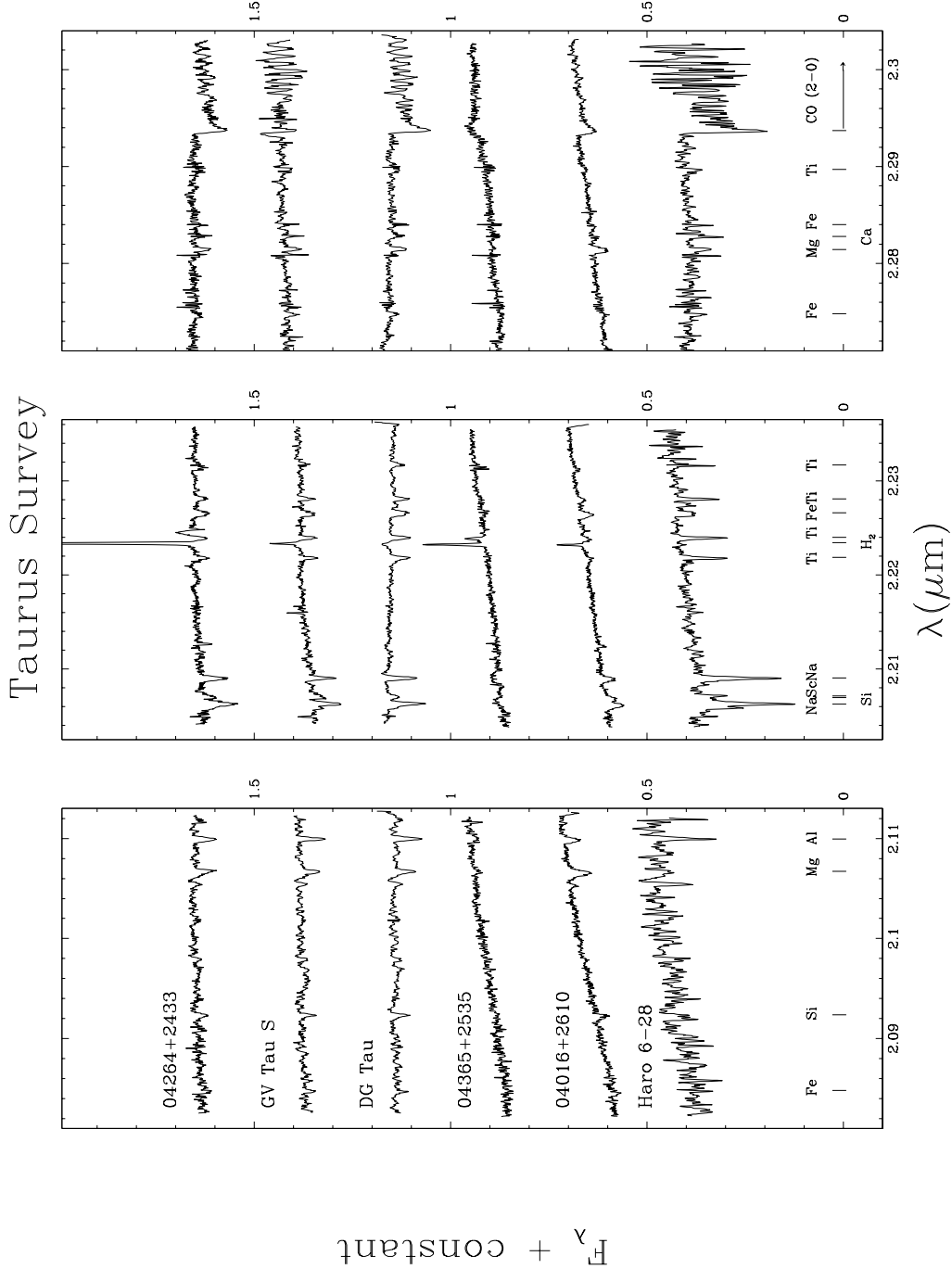}
\caption[Taurus Spectra1]
{\label{fig-all.spectra1}
High resolution (R $\simeq$ 18,000) spectra of protostars in
Taurus-Auriga are shown in three echelle orders, which contain the
photospheric absorption line regions (Mg/Al, Na, and $^{12}$CO
intervals, bottom to top, respectively) that we use to derive physical
properties using spectral fits to synthetic templates.  At this
resolution, the intrinsic line widths of K-band photospheric lines are
fully resolved allowing precise fits to their line shapes.}

\end{figure}


\clearpage




\clearpage
\oddsidemargin=-1cm
\tabletypesize{\scriptsize}

\begin{deluxetable}{lcrrcrrc}
\tablecaption{Journal of Observations \label{tbl-1}}
\tabletypesize{\scriptsize}
\tablewidth{0 pt}
\tablehead{
\colhead{Source} & 
\colhead{Region or} & 
\colhead{$\alpha$(J2000)} & 
\colhead{$\delta$(J2000)}  & 
\colhead{UT Date} & 
\colhead{Int. Time} & 
\colhead{S/N} & 
\colhead{Photospheric}\\ 
& 
\colhead{Sp Type} & 
\colhead{hh mm ss.ss} & 
\colhead{$\arcdeg$~$\arcmin$~$\arcsec$}  &
& 
\colhead{(minutes)} & 
& 
\colhead{Lines?}}
 
\startdata
 
04016+2610      &       Tau         &       4       4       43.10   & 26      18      57.9    &       4-Nov-01        &       25.0    &       240 &       yes\\
04108+2803B     &       Tau         &       4       13      54.89   & 28      11      30.5    &       6-Nov-01        &       20.0    &       110 &       yes\\
04158+2805      &       Tau         &       4       18      58.17   & 28      12      23.7    &       6-Nov-01        &       40.0    &       70 &       yes\\
04181+2655      &       Tau         &       4       21      10.94   & 27      2       5.9    &       6-Nov-01        &       30.0    &       120 &       yes\\
DG~Tau          &       Tau         &       4       27       4.77   & 26      6       17.4    &       6-Nov-01        &       4.0     &       240 &       yes\\
GV~Tau~S        &       Tau         &       4       29      23.66   & 24      33      2.0     &       6-Nov-01        &       16.0    &       220 &       yes\\
04264+2433       &      Tau         &       4       29      30.01   & 24      39      55.6    &       6-Nov-01        &       48.0    &       190 &       yes\\
L1551~IRS5      &       Tau         &       4       31      34.10   & 8       8       8.0     &       6-Nov-01        &       20.0    &       100 &       yes\\
04295+2251      &       Tau         &       4       32      32.07   & 22      57      30.3    &       5-Nov-01        &       33.3    &       205 &       yes\\
Haro~6-28       &       Tau         &       4       35      56.77   & 22      54      36.5    &       5-Nov-01        &       20.0    &       110 &       yes\\
04361+2547      &       Tau         &       4       39      13.43   & 25      53      19.2    &       5-Nov-01        &       44.0    &       230 &       no\\
04365+2535      &       Tau         &       4       39      34.99   & 25      41      46.7    &       4-Nov-01        &       60.0    &       210 &       no\\
04489+3042      &       Tau         &       4       52      6.87    & 30      47      17.2    &       4-Nov-01        &       16.7    &       120 &       yes\\
\\
CRBR~12          &       Oph        &       16      26      17.23   & -24     23      44.6    &       20-Jun-03       &       50.0    &       50 &       yes\\
CRBR~85          &       Oph        &       16      26      17.23   & -24     23      44.6    &       20-Jun-03       &       120.0   &       15 &       no\tablenotemark{a}\\
GSS~30           &       Oph        &       16      26      21.42   & -24     23      6.3     &       8-Jul-01        &       30.0    &       395 &       no\\
GY~21            &       Oph        &       16      26      23.56   & -24     24      38.2    &       10-Jul-01    &       16.7    & 170     &       yes\\
162636          &       Oph        &       16      26      36.90   & -24     15      54.0    &       10-Jul-01       &       15.0    &       100 &       yes\\
GY~91            &       Oph        &       16      26      40.43   & -24     27      15.1    &       21-Jun-03       &       40.0    &       55 &       yes\\
WL~12\tablenotemark{b}            &       Oph        &       16      26      44.10   & -24     34      47.8    &       7-Jul-01,8-Jul-01       &       87.3    & 215     &       yes\\
WL~1             &       Oph        &       16      27      3.98    & -24     28      27.5    &       30-May-00       &       30.0    &       35 &       yes\\
WL~17            &       Oph        &       16      27      6.69    & -24     38      15.3    &       10-Jul-01       &       30.0    &       140 &       yes\\
GY~224           &       Oph        &       16      27      11.25   & -24     40      47.0    &       10-Jul-01       &       50.0    &       150 &       yes\\
WL~19            &       Oph        &       16      27      11.60   & -24     38      32.0    &       30-May-00       &       20.0    &       85 &       yes\\
WL~3             &       Oph        &       16      27      19.20   & -24     28      40.5    &       10-Jul-01       &       40.0    &       95 &       yes\\
WL~6             &       Oph        &       16      27      21.62   & -24     29      51.3    &       19-Jun-03       &       40.0    &       150 &       no\\
IRS~43           &       Oph        &       16      27      26.96   & -24     40      50.0    &       7-Jul-01        &       73.3    &       330 &       yes\\
YLW~16A\tablenotemark{b}          &       Oph        &       16      27      27.84   & -24     39      31.9    &       8-Jul-01,9-Jul-01       &       38.0    & 170     &       yes\\
YLW~16B          &       Oph        &       16      27      29.29   & -24     39      15.8    &       30-May-00       &       20.0    &       180 &       no\\
VSSG~17          &       Oph        &       16      27      30.39   & -24     27      43.7    &       10-Jul-01       &       6.0     &       160 &       yes\\
IRS~51\tablenotemark{c}           &       Oph        &       16      27      39.53   & -24     43      14.1    &       8-Jul-01,20-Jun-03      &       50.0    & 330     &       yes\\
IRS~63           &       Oph        &       16      31      35.53   & -24     1       28.3    &       30-May-00       &       30.0    &       140 &       yes\\
IRS~67\tablenotemark{b}           &       Oph        &       16      32      1.05    & -24     56      44.6    &       7-Jul-01,8-Jul-01       &       80.0    & 330     &       yes\\
\\
EC~38            &       Ser        &       18      29      49.50   & 1       17      7.0     &       20-Jun-03       &       90.0    &       70 &       yes\\
EC~53\tablenotemark{b}            &       Ser        &       18      29      51.20   & 1       16      42.0    &       7-Jul-01,19-Jun-03      &       104.0   & 120     &       yes\\
SVS~2            &       Ser        &       18      29      56.79   & 1       14      46.1    &       29-May-00       &       12.0    &       150 &       no\\
EC~89            &       Ser        &       18      29      57.50   & 1       12      59.0    &       8-Jul-01        &       40       &       40 &       no\tablenotemark{a}\\
EC~88            &       Ser        &       18      29      57.50   & 1       12      59.0    &       21-Jun-03       &       60.0    &       60 &       no\\
EC~91            &       Ser        &       18      29      57.70   & 1       12      28.0    &       20-Jun-03       &       60.0    &       50 &       yes\\
EC~92            &       Ser        &       18      29      57.70   & 1       12      52.0    &       7-Jul-01        &       46.0    &       195 &       yes\\
SVS~20A          &       Ser        &       18      29      57.77   & 1       14      7.2     &       30-May-00       &       4.0     &       120 &       yes\\
EC~94            &       Ser        &       18      29      57.80   & 1       12      37.0    &       19-Jun-03       &       60.0    &       80 &       yes\\
EC~95            &       Ser        &       18      29      57.90   & 1       12      47.2    &       30-May-00       &       20.0    &       140 &       yes\\
EC~103           &       Ser        &       18      29      58.70   & 1       14      26.0    &       8-Jul-01        &       40.0    &       145 &       no\\
EC~125           &       Ser        &       18      30      1.90    & 1       14      1.0     &       21-Jun-03       &       30.0    &       45 &       yes\\
EC~129           &       Ser        &       18      30      2.80    & 1       12      28.0    &       8-Jul-01        &       38.7    &       210 &       yes\\
\\
TS~13.1          &       CrA        &       19      1       41.40   & -36     58      34.0    &       29-May-00       &       8.0     &       180 &       yes\\
RCrA~IRS5       &       CrA        &       19      1       48.00   & -36     57      19.0    &       21-Jun-03       &       10.0    &       70 &       yes\\
TS~2.6           &       CrA        &       19      1       49.78   & -36     58      12.0    &       29-May-00       &       4.0     &       190 &       no\\
RCrA~IRS9       &       CrA        &       19      1       53.10   & -36     57      4.0     &       21-Jun-03       &       53.3    &       90 &       yes\\
\\
03260+3111A     &       Per         &       3       29      10.49   & 31      21      51.2    &       5-Nov-01        &       6.0     &       205 &       yes\\
03260+3111B     &       Per         &       3       29      10.49   & 31      21      51.2    &       5-Nov-01        &       20.0    &       70 &       yes\\
\\
MK Standards\\
\\
GL~15A           &       M2~V        &       0       18      18.70   & 44      1       23.3    &       29-May-00       &       1.0     &       110 &       yes\\
GL~28            &       K2~V        &       0       40      48.40   & 40      11      14.7    &       29-May-00       &       2.0     &       115 &       yes\\
GL~338B          &       K7~V        &       9       14      26.50   & 52      41      15.9    &       30-May-00       &       0.7     &       200 &       yes\\
GL~338A          &       M0~V        &       9       14      28.70   & 52      41      15.3    &       29-May-00       &       2.0     &       185 &       yes\\
GJ~402           &       M4~V        &       10      50      52.40   & 6       48      31.6    &       30-May-00       &       5.0     &       110 &       yes\\
GJ~406           &       M6~V        &       10      56      29.10   & 7       0       56.1    &       30-May-00       &       4.0     &       75 &       yes\\
HR~4358          &       K3~III      &       11      14      1.80    & 8       3       38.0    &       30-May-00       &       1.3     &       180 &       yes\\
HR~4496          &       G8~V        &       11      41      3.20    & 34      12      5.7     &       29-May-00       &       1.0     &       215 &       yes\\
HR~5150          &       M1.5~III    &       13      41      36.70   & -8      42      11.0    &       30-May-00       &       0.2     &       130 &       yes\\
LHS~2924        &       M9~V        &       14      28      43.22   & 33      10      35.3    &       20-Jun-03       &       30.0    &       300 &       yes\\
GJ~569B          &       M8.5~V      &       14      54      28.97   & 16      6       4.8     &       19-Jun-03       &       16.0    &       265 &       yes\\
GJ~569A          &       M3~V        &       14      54      29.31   & 16      6       3.4     &       19-Jun-03       &       0.7     &       265 &       yes\\
HD~131976        &       M1~V        &       14      57      26.79   & -21     24      26.8    &       21-Jun-03       &       0.4     &       450 &       yes\\
HR~5568          &       K4~V        &       14      57      27.90   & -21     24      56.0    &       29-May-00       &       1.0     &       240 &       yes\\
VB~8             &       M7~V        &       16      55      35.16   & -8      23      45.4    &       21-Jun-03       &       6.0     &       260 &       yes\\
HD~157881        &       K7~V        &       17      25      45.10   & 2       6       37.0    &       21-Jun-03       &       0.6     &       350 &       yes\\
HD~166620        &       K2~V        &       18      9       37.32   & 38      27      26.3    &       21-Jun-03       &       0.5     &       315 &       yes\\
HD~168387        &       K2~III      &       18      19      9.53    & 7       15      35.1    &       21-Jun-03       &       0.1     &       250 &       yes\\
HD~175225        &       K1~IV       &       18      51      34.88   & 52      58      30.0    &       21-Jun-03       &       0.2     &       220 &       yes\\
VB~10            &       M8~V        &       19      16      57.53   & 5       8       55.5    &       20-Jun-03       &       10.0    &       295 &       yes\\
HD~182572        &       G8~IV       &       19      24      58.22   & 11      56      40.2    &       20-Jun-03       &       0.1     &       360 &       yes\\
HD~184489        &       M0~V        &       19      34      39.96   & 4       34      58.1    &       20-Jun-03       &       0.8     &       400 &       yes\\
GJ~1245A         &       M5.5~V      &       19      53      54.58   & 44      24      49.8    &       19-Jun-03       &       2.7     &       360 &       yes\\
GJ~791.2         &       M4.5~V      &       20      29      48.57   & 9       41      19.6    &       20-Jun-03       &       3.0     &       255 &       yes\\
GJ~806           &       M2~V        &       20      45      4.24    & 44      29      57.6    &       19-Jun-03       &       2.0     &       175 &       yes\\
HD~201091        &       K5~V        &       21      6       55.19   & 38      45      9.3     &       21-Jun-03       &       0.1     &       250 &       yes\\
GJ~4281          &       M6.5~V      &       22      28      54.52   & -13     25      25.0    &       21-Jun-03       &       16.0    &       240 &       yes\\
HD~219134        &       K3~V        &       23      13      17.87   & 57      10      7.1     &       21-Jun-03       &       0.4     &       433 &       yes\\
\enddata
\tablenotetext{a}{CO and Na line absorption is marginally detected in this source, but at too low a signal to noise needed for us to apply our standard
fitting technique.
}
\tablenotetext{b}{Source was observed on multiple nights.  The final spectrum
is the noise weighted sum of spectra obtained on each individual night.  No shifting of the spectra was required as radial velocity differences were less than 
1 pixel ($\sim$4.5 km~s$^{-1}$).
}
\tablenotetext{c}{The final spectrum
is the noise weighted sum of spectra obtained on both nights.
The 8 July observation exhibits slightly broader and 
asymmetric lines suggesting the presence of a cooler ($<$ 3900 K) 
binary companion with a marginal velocity separation ($\le$ 2 resolution
elements). 
}
\end{deluxetable}

\clearpage


\begin{deluxetable}{lcccccc}
\tabletypesize{\scriptsize}
\tablecaption{Census of Spectral Line Detections\label{tbl-2}}
\tablewidth{0 pt}
\tablehead{
\colhead{Spectral Lines} &
\colhead{Tau} & 
\colhead{Oph} & 
\colhead{Ser} & 
\colhead{CrA} & 
\colhead{Per} &
\colhead{Total}}

\startdata

Mg/Al, Na, or $^{12}$CO (absorption)\tablenotemark{a}     &  11/13  &  16/20  & 9/13   &  3/4  & 2/2  &  41/52\\
Br~$\gamma$ (absorption)\tablenotemark{b}    &  1/13  &  2/20  & 6/13   &  0/4  & 1/2  &  10/52\\
Br~$\gamma$ (emission)\tablenotemark{b}    &  9/13  &  13/20  & 9/13   &  3/4  & 0/2  &  34/52\\
$^{12}$CO (emission)\tablenotemark{a}   &  2/13   &  1/20   & 3/13   &  2/4  & 0/2  &  8/52\\
H$_2$ 1-0 S(0)  (emission)     &  10/13  &  8/20  &  3/13   &  1/4  & 1/2  &  23/52\\

\enddata

\tablenotetext{a}{CO emission and absorption components were
present in 2 sources (EC~129 \& SVS~20A) and are included here.}
\tablenotetext{b}{Six sources showed both Br~$\gamma$ emission and
absorption, and are included in here.}

\end{deluxetable}



\clearpage

\begin{deluxetable}{lrcccccrccc}
\tabletypesize{\scriptsize}
\tablecaption{Derived YSO properties \label{tbl-3}}
\tablewidth{0pt}
\tablehead{
\colhead{Source} & 
\colhead{T$_{\rm eff}$} & 
\colhead{error} & 
\colhead{$\log~g$}  & 
\colhead{error} & 
\colhead{$v\sin~i$} & 
\colhead{error} & 
\colhead{r$_{\rm K}$} & 
\colhead{error} & 
\colhead{v$_{\rm lsr}$\tablenotemark{a}} & 
\colhead{L$_*$\tablenotemark{b}} \\ 
& 
\colhead{(K)} &  
& 
\colhead{(cm s$^{-2}$)} & 
& 
\colhead{(km~s$^{-1}$)} &  
& 
& 
& 
\colhead{(km~s$^{-1}$)} & 
\colhead{(L/L$_{\odot}$)}}

\startdata

04016+2610	&	4500	&	$\pm$68	&	$\le$3.5	&	$\pm$0.05	&	46	&	$\pm$2.7	&	0.9	&	$\pm$0.02	&	-12.8	&	4.9\\
04108+2803B	&	3500	&	$\pm$22	&	$\le$3.5	&	$\pm$0.03	&	14	&	$\pm$3.3	&	1.1	&	$\pm$0.08	&	5.4	&	0.4\\
04158+2805	&	3500	&	+213, -124	&	3.8	&	 +0.40, -0.11 	&	26	&	$\pm$1.2	&	1.3	&	 +0.21, -0.19 	&	5.9	&	0.1\\
04181+2655	&	4000	&	+202, -301	&	4.2	&	 +0.30, -0.50 	&	33	&	 +1.3, -0.9 	&	0.2	&	$\pm$0.004	&	5.6	&	1.8\\
DG~Tau	&	4000	&	 +105, -33 	&	4.0	&	 +0.12, -0.06 	&	24	&	$\pm$2.0	&	2.0	&	$\pm$0.10	&	7.1	&	5.5\\
GV~Tau~S        &	4500	&	$\pm$25	&	4.4	&	$\pm$0.06	&	27	&	$\pm$1.6	&	1.8	&	$\pm$0.003	&	-7.2	&	3.3\\
04264+2433	&	4000	&	$\pm$41	&	3.9	&	$\pm$0.12	&	32	&	$\pm$2.4	&	1.4	&	$\pm$0.03	&	4.3	&	0.9\\
L1551~IRS5	&	4800	&	$\pm$154	&	4.0	&	$\pm$0.14	&	31	&	$\pm$2.4	&	1.0	&	$\pm$0.30	&	1.2	&	2.6\\
04295+2251	&	3400	&	$\pm$40	&	$\le$3.5	&	$\pm$0.02	&	51	&	$\pm$1.9	&	0.3	&	$\pm$0.01	&	3.7	&	1.2\\
Haro~6-28	&	3400	&	 +35, -106 	&	4.1	&	 +0.06, -0.31 	&	17	&	 +2.2, -2.4 	&	0.6	&	 +0.11, -0.04 	&	3.5	&	0.4\\
04489+3042	&	3100	&	$\pm$19	&	$\le$3.5	&	$\pm$0.07	&	17	&	$\pm$2.2	&	1.4	&	$\pm$0.11	&	-17.5	&	0.3\\
\\
CRBR~12	&	3600	&	 +89, -313 	&	4.5	&	 +0.09, -0.71 	&	40	&	$\pm$3.7	&	2.0	&	 +0.64, -0.57 	&	-8.5	&	1.0\\
GY~21	&	3900	&	$\pm$37	&	3.8	&	$\pm$0.09	&	24	&	$\pm$2.4	&	1.4	&	$\pm$0.03	&	1.5	&	0.8\\
162636	&	3700	&	$\pm$56	&	4.3	&	$\pm$0.09	&	27	&	$\pm$4.7	&	1.9	&	$\pm$0.06	&	3.7	&	0.4\\
GY~91	&	3300	&	$\pm$30	&	3.7	&	$\pm$0.03	&	12	&	$\pm$0.9	&	0.3	&	$\pm$0.01	&	3.5	&	1.7\\
WL~12	&	4000	&	 +403, -46 	&	3.6	&	 +0.71, -0.12 	&	35	&	$\pm$3.4	&	2.8	&	 +0.71, -0.12 	&	1.2	&	2.1\\
WL~1	&	3300	&	$\pm$12	&	$\le$3.5	&	$\pm$0.02	&	14	&	$\pm$0.8	&	0.8	&	$\pm$0.004	&	-12.7	&	0.7\\
WL~17	&	3400	&	 +203, -34 	&	3.7	&	 +0.4, -0.05 	&	12	&	 +4.3, -4.4 	&	3.9	&	 +0.77, -0.80 	&	4.4	&	1.8\\
GY~224	&	3500	&	 +306, -116 	&	3.7	&	 +0.50, -0.21 	&	10	&	$\pm$5.1	&	4.4	&	$\pm$0.75	&	2.2	&	0.7\\
WL~19	&	4200	&	 +143, -246 	&	4.6	&	 +0.19, -0.44 	&	19	&	 +3.5, -3.6 	&	3.7	&	 +0.31, -0.37 	&	-16.6	&	10.3\\
WL~3	&	3300	&	 +202, -28 	&	3.7	&	 +0.41, -0.08 	&	37	&	$\pm$2.1	&	0.5	&	 +0.04, -0.11 	&	-2.3	&	2.3\\
IRS~43	&	4400	&	 +118, -405 	&	4.2	&	 +0.32, -0.61 	&	48	&	 +3.5, -3.9 	&	1.8	&	 +0.11, -0.20 	&	1.8	&	19.0\\
YLW~16A	&	4400	&	 +62, -118 	&	4.5	&	 +0.10, -0.22 	&	29	&	$\pm$3.5	&	2.2	&	 +0.23, -0.25 	&	3.7	&	9.8\\
VSSG~17	&	3500	&	$\pm$42	&	$\le$3.5	&	$\pm$0.08	&	43	&	$\pm$1.7	&	0.5	&	$\pm$0.005	&	1.0	&	9.7\\
IRS~51	&	4200	&	 +208, -116 	&	3.7	&	 +0.30, -0.11 	&	44	&	$\pm$2.8	&	1.8	&	 +0.10, -0.01 	&	9.1	&	21.7\\
IRS~63	&	4200	&	 +144, -103 	&	$\le$3.5	&	 +0.13, -0.08 	&	45	&	 +3.1, -2.9 	&	1.7	&	 +0.32, -0.34 	&	-26.5	&	3.8\\
IRS~67	&	4300	&	$\pm$91	&	$\le$3.5	&	$\pm$0.10	&	55	&	$\pm$3.0	&	1.0	&	$\pm$0.02	&	-0.7	&	11.6\\
\\
EC~38	&	3400	&	 +107, -37 	&	$\le$3.5	&	 +0.22, -0.10 	&	34	&	 +2.5, -2.7 	&	0.9	&	 +0.10, -0.006 	&	10.8	&	1.9\\
EC~53	&	3400	&	 +304, -47 	&	$\le$3.5	&	 +0.41, -0.08 	&	24	&	$\pm$4.3	&	1.8	&	 +0.40, -0.26 	&	4.3	&	0.4\\
EC~91	&	3600	&	$\pm$74	&	3.8	&	$\pm$0.14	&	17	&	$\pm$0.9	&	1.1	&	$\pm$0.07	&	4.8	&	1.0\\
EC~92	&	4200	&	$\pm$17	&	$\le$3.5	&	$\pm$0.02	&	47	&	$\pm$0.9	&	0.0	&	+0.002	&	9.4	&	6.5\\
SVS~20A	&	5900	&	 +205, -228 	&	$\le$3.5	&	$\pm$0.23	&	52	&	 +6.0, -6.1 	&	0.7	&	 +0.1, -0.007 	&	-8.3	&	55.6\\
EC~94	&	4000	&	 +301, - 24 	&	3.6	&	 +0.30, -0.05 	&	36	&	$\pm$1.6	&	0.5	&	$\pm$0.004	&	5.8	&	4.3\\
EC~95	&	4400	&	 +115, -57 	&	$\le$3.5	&	 +0.10, -0.03 	&	56	&	$\pm$1.1	&	0.1	&	$\pm$0.002	&	-14.5	&	22.8\\
EC~125	&	$\le$3000	&	$\pm$26	&	4.0	&	$\pm$0.26	&	30	&	$\pm$2.4	&	0.8	&	$\pm$0.09	&	5.7	&	0.2\\
EC~129	&	4400	&	$\pm$74	&	$\le$3.5	&	$\pm$0.06	&	35	&	$\pm$2.2	&	0.6	&	$\pm$0.004	&	10.0	&	5.2\\
\\
TS~13.1	&	4400	&	 +69, -212 	&	$\ge$4.5	&	 +0.14, -0.42 	&	34	&	 +2.6, -2.8 	&	2.7	&	 +0.03, -0.2 	&	14.5	&	...\\
RCrA~IRS5	&	3700	&	 +203, -107 	&	3.7	&	 +0.31, -0.21 	&	36	&	$\pm$1.1	&	0.3	&	$\pm$0.01	&	1.6	&	...\\
RCrA~IRS9	&	3800	&	 +100, -9 	&	4.1	&	 +0.2, -0.01 	&	5	&	 +1.9, -1.6 	&	0.0	&	$\pm$0.01	&	5.3	&	...\\
\\
03260+3111A	&	5600	&	$\pm$112	&	$\le$3.5	&	$\pm$0.13	&	76	&	$\pm$4.6	&	0.0	&	$\pm$0.01	&	6.2	&	...\\
03260+3111B	&	3400	&	 +39, -204 	&	3.9	&	 +0.05, -0.30 	&	23	&	$\pm$1.6	&	0.3	&	$\pm$0.04	&	3.9	&	...\\
\enddata

\tablenotetext{a}{A systematic shift of +1.8 km~s$^{-1}$ has been applied to these values, consistent with our radial velocity measurements measured in MK standards with published values (Fig. \ref{fig-radvel.offset}).}
\tablenotetext{b}{Stellar luminosities are based on derived K-band extinctions which have been elevated by 0.88 magnitudes to account for the effects of scattered light (see $\S$ \ref{sec-photlums}).}

\end{deluxetable}






\end{document}